\documentclass[prd,aps,showpacs,tightenlines,notitlepage,nofootinbib,superscriptaddress,groupedaddress]{revtex4-1}
\usepackage{mathrsfs}
\usepackage{amsmath}
\usepackage{amsfonts}
\usepackage{array}
\usepackage{verbatim}
\usepackage{epsfig}
\usepackage{graphicx} 
\usepackage[usenames, dvipsnames]{color}
\usepackage{marginnote}
\usepackage{slashed}
\usepackage{bm}

\newcounter{mynote}

\begin{document}
\title{Structure functions at small x from world-lines. I: Unpolarized distributions}
\author{Andrey Tarasov}
\author{Raju Venugopalan}

\affiliation{Physics Department, Brookhaven National Laboratory,
Bldg. 510A, Upton, NY 11973, U.S.A.}

\begin{abstract}
The world-line representation of quantum field theory is a powerful framework for the computation of perturbative multi-leg Feynman amplitudes. In particular, in gauge theories, it provides an efficient way, via point particle Grassmann functional integrals, to compute spinor and color traces in these amplitudes. Further, semi-classical approximations to quantum mechanical world-line trajectories  provide useful intuition in a wide range of dynamical problems. We develop here the world-line approach to compute deeply inelastic structure functions in the small $x$ Regge limit of QCD. In a shockwave approximation valid in this limit, we show how one recovers the well-known dipole model for unpolarized structure functions. In a follow-up work, we will discuss the world-line computation of polarized structure functions at small $x$. 
\end{abstract}

\maketitle

\section{Introduction}
In quantum field theory, one and higher loop effective actions can be rewritten as quantum mechanical point particle path integrals~\cite{Schwinger:1951nm}, with internal degrees such as spin and color expressed as Grassmann degrees of freedom~\cite{Berezin:1976eg,Balachandran:1976ya,Balachandran:1977ub,Barducci:1976xq,Barducci:1982yw,Brink:1976uf,Barut:1988ee}. An elegant feature of the formalism is that the point particle degrees of freedom can be thought of as world-lines tracing paths in the presence of background fields. Indeed, the pseudo-classical Bargmann-Michel-Telegdi (BMT) equation~\cite{Bargmann:1959gz} for spinning particles in background fields, and the Wong equations~\cite{Wong:1970fu} for their color charge counterparts, are straightforwardly obtained by taking the saddle points of the respective QED and QCD one-loop world-line path integrals; in gravity, the world-line counterparts are the Papapetrou equations~\cite{Papapetrou:1951pa,Barducci:1976xq}. Because of the semi-classical intuition provided by world-lines, they are a powerful tool in first principles derivations of phase space distributions~\cite{Berezin:1976eg}. This has been exploited in developing kinetic descriptions in QED and QCD at finite temperature and density~\cite{JalilianMarian:1999xt,Mueller:2017lzw,Mueller:2017arw,Mueller:2019gjj}. 

In QED, the world-line formalism is one loop exact; in QCD, it can be derived in a systematic perturbative expansion of the effective action. 
Indeed, the world-line framework provides an elegant and efficient way to compute Feynman amplitudes~\cite{Strassler:1992zr,DHoker:1995uyv,DHoker:1995aat,Mondragon:1995va,Mondragon:1995ab,Hernandez:2008db,Schubert:2001he,Bastianelli:2006rx,Bastianelli:2013pta,Corradini:2015tik}. In the case of QCD for instance, there is an equivalence~\cite{Strassler:1992zr} between this approach and the rules derived by Bern, Dixon, Dunbar and Kosower~\cite{Bern:1991aq,Bern:1994cg} from string theory to compute one-loop gauge theory amplitudes. A further important development has been the application of world-line techniques to describe soft gluon resummation and exponentiation in QCD. The exponentiation of a class of soft graphs called 
``webs"~\cite{Berger:2003zh} finds an elegant description in the world-line approach~\cite{Laenen:2008gt,Gardi:2010rn,Laenen:2010uz}. In fact, this exponentiation can be understood as the cloud of soft radiation dressing the trajectories of hard world-line charges and is deeply related to the infrared finiteness of S-matrix elements in high energy scattering~\cite{Kulish:1970ut,Catani:1985ta,Catani:1987sp,Giavarini:1987ts}.

In this work, we will apply the world-line formalism to deeply inelastic scattering (DIS) in the Regge limit of fixed $Q^2 \gg \Lambda_{\rm QCD}^2$ and  Bjorken $x\rightarrow 0$. In this limit, soft radiation is dominated by large logs $\alpha_s \ln(1/x)\sim O(1)$, where $\alpha_s$ is the QCD coupling. The resummation of these leading logarithms in $x$ (LLx), to all orders in perturbation theory, is described\footnote{For a derivation of the BFKL equation in a framework where the motion of hard color charged particles is described by Wong's equations, see Ref.~\cite{JalilianMarian:2000ad}.} by the BFKL equation~\cite{Kuraev:1977fs,Balitsky:1978ic};  the LLx resummation was subsequently extended to next-to-leading-logarithmic accuracy~\cite{Fadin:1998py}. This small $x$ resummation leads to a rapid growth of parton distributions; in fact, for each $Q^2$, there is a value of $x$ below which the phase space density of partons on the light front has maximal occupancy of order $1/\alpha_s$. This phenomenon is called gluon saturation and the emergent scale $Q_s(x)$ where it occurs is called the saturation scale~\cite{Gribov:1984tu,Mueller:1985wy}. In the Regge limit, $Q_s \gg \Lambda_{\rm QCD}$; the coupling is weak, with $\alpha_s\equiv\alpha_s(Q_s)$. The dynamics of gluon saturation is however nonperturbative due to the large gluon occupancy. 

The physics of this weakly coupled albeit nonperturbative gluon saturation regime of QCD is captured by an effective field theory (EFT), the Color Glass Condensate (CGC)~\cite{Iancu:2003xm,Gelis:2010nm,Kovchegov:2012mbw,Blaizot:2016qgz} where the degrees of freedom are static color sources at large $x$ and dynamical fields at small $x$. As first noted in \cite{McLerran:1993ni, McLerran:1993ka, McLerran:1994vd}, because of the large gluon occupancy, the CGC is a classical EFT. Remarkably, Wilsonian renormalization group (RG) computations  show that the structure of the classical field reproduces itself with decreasing $x$, while the distribution of color sources is modified with each step in $x$; the corresponding RG equations for the sources are called the JIMWLK equations \cite{JalilianMarian:1996xn,JalilianMarian:1997gr,JalilianMarian:1997dw,Iancu:2000hn,Ferreiro:2001qy}. 

In DIS at small $x$, the problem can be formulated equivalently as the fluctuation of the virtual photon into a quark-antiquark pair that scatters off the ``shockwave" classical background field of a nucleus and is color rotated in the process~\cite{Nikolaev:1990ja,Mueller:1989st}. In general, the RG that emerges can be understood as generating a hierarchy of equations for products of light-like Wilson lines (and their Hermitean conjugates) representing the color rotation of the quark, anti-quark and additional partons from the projectile in the shockwave background. The resulting Balitsky hierarchy~\cite{Balitsky:1995ub} is identical~\cite{Weigert:2000gi} to that generated by the JIMWLK RG equation. For the RG evolution of the quark-antiquark dipole,  one obtains a closed form expression, the Balitsky-Kovchegov (BK) equation, in a mean field large $N_c$ approximation~\cite{Balitsky:1995ub,Kovchegov:1999yj}. 

In this paper, we will provide a first demonstration of the power of the world-line approach in computations at small $x$. While the world-line approach is equivalent to the standard formulation of QCD, it provides an alternative representation which is useful not only for perturbative calculations but also for non-perturbative problems in QCD. From the perspective of perturbative calculations,  the world-line approach at small $x$ has several attractive features relative to the standard formalism of Feynman diagrams. Firstly, the world-line framework is formulated in coordinate space, which allows one to introduce the ``shock wave" representation of small $x$ gauge fields as an instantaneous interaction of the world-line with the background gauge field. Further, the interaction terms of the world-line action immediately lead to gauge invariant expressions which are constructed from the field strength tensor and Wilson lines of the background field.  In the standard perturbative approach, such a formulation is a consequence of the resummation of a large number of diagrams from different orders of expansion in the background field, and reorganization of these diagrams into gauge invariant objects. The realization of this operation in the world-line approach is more transparent, arising directly from the structure of the world-line Hamiltonian. This is especially important at small $x$ because all twists in the expansion of background field are equally important.

Another interesting feature of the world-line approach is the explicit separation between fermionic and bosonic degrees of freedom already at the level of the world-line action.  This makes the world-line approach very powerful for study of spin effects. In  Feynman diagram calculations, these degrees of freedom are mixed. In particular, in this paper, we derive the structure of the interaction current of the world-line with the shock-wave background and observe that it has a spin-dependent term. This term in the current is specified by the shock-wave approximation. In the derivation, we used the properties of Grassmann variables that describe the transition of polarization from the first non-trivial correction to the background field to the fermion degrees of freedom. While similar structures have been observed before \cite{Kovchegov:2015pbl,Kovchegov:2016weo,Kovchegov:2018znm,Chirilli:2018kkw} in the Feynman diagram computations at small $x$, we believe that our result is the first derivation, valid in all orders of perturbation theory, that demonstrates that the structure of the spin-dependent interaction at small $x$ is a direct consequence of the instantaneous nature of the interaction with the background.  
In a forthcoming publication \cite{inprep:2019} (paper II in this series), we will use the structure of the spin-dependent current to derive the form of the structure function $g_1$ at small $x$ and introduce the notion of the polarized dipole. 

Finally, the world-line approach is ideally suited to address the subtle issues regarding the role of non-perturbative effects at small $x$~\cite{Mueller:2017lzw,Mueller:2017arw}. In particular, it can potentially provide insight into  the role of the chiral anomaly in polarized DIS at small $x$~\cite{Carlitz:1988ab,Altarelli:1988nr,Jaffe:1989jz}. Another interesting possibility is that the world-line approach may allow one to use a quantum computer to determine structure functions at small $x$ in arbitrary non-perturbative background fields. A paper outlining the power of the world-line framework in quantum computation is in preparation~\cite{inprepQC:2019}.

In this paper, for a first application of the world-line formalism to DIS at small $x$,  we will only consider unpolarized structure functions and shall derive the dipole model for these structure functions from first principles. As we will describe, it is sufficient for our purposes to treat the CGC shockwave field as the background field in this formalism. The RG evolution is contained in the evolution of the classical shockwave with decreasing $x$ or increasing center-of-mass energy. Our approach is similar in spirit to that developed in \cite{McLerran:1998nk,Venugopalan:1999wu} albeit the latter is couched in the language of Feynman diagrams. We will establish a primer between the two descriptions. As noted, in a follow-up paper II, we will use the world-line techniques adapted here to describe the Regge limit to study the corresponding spinning dipole model for polarized parton distributions~\cite{inprep:2019}. 

The paper is organized as follows. In the next section, we will begin with some preliminaries on definitions of structure functions and relate these to the imaginary part of the time ordered product of electromagnetic currents. These currents are obtained from varying the effective action with respect to the electromagnetic background field. This is the starting point for the application of the world-line formalism to DIS. In Section III, we will provide an introduction to the formalism for the general reader. We start with scalar QED and then generalize to full QED. This provides us with the ingredients to compute the vacuum polarization tensor. The polarization tensor in the gluon shockwave background field is discussed in Section IV and the dipole model derived in Section V. We end with a summary and conclusions. 

For the interested reader, the Appendices provide useful and, in some instances, novel information. In Appendix A, we introduce the method of Grassmann coherent states to derive the world-line path integral for spinning particles. Thus one effectively replaces spinor traces with quantum mechanical Grassmann integrals. A similar procedure can be followed to compute color traces. We do not discuss this here but note some of the relevant references in this regard to be \cite{Barducci:1980xk,DHoker:1995uyv,DHoker:1995aat,Mondragon:1995va,Mondragon:1995ab,Edwards:2017nvs,Mueller:2017arw,Mueller:2019gjj}. Appendices B and C discuss the computation of scalar and Grassmann functional integrals respectively. An excellent review for this purpose can be found in \cite{Schubert:2001he}. Specifically, we here make use of the expressions for the boson and fermion world-line Green's functions on a closed loop discussed previously in \cite{Strassler:1992zr}. A key observation is that the semi-classical expressions for the world-line functional integrals are exact in the shockwave approximation. In Appendix D, we reformulate the coordinate space expressions for world-line functional integrals in momentum space. This has the advantage that a simple mnemonic can be used to describe one loop integrals with an arbitrary number of external currents. That is not the case for the coordinate space equivalent. Further, while the representation of proper time integrals is long known to be equivalent to the Feynman parameter representation of propagators, we highlight subtle features of the world-line representation. Finally, in Appendix E, we discuss the dictionary between Feynman diagrams and world-line path integrals. We discuss some subtleties in matching spinor currents in the two approaches that will be relevant in future discussions of polarized parton distributions.

\section{Deep inelastic scattering, structure functions and small $x$}
\label{section:II}
The subject of our interest is inclusive deeply inelastic lepton-hadron scattering (DIS), summarized by the expression 
\begin{eqnarray}
l(l) + N(P)\to l(l') + X,
\end{eqnarray}
where the interaction between the lepton ($l$) and the hadron ($N$) is viewed as the exchange of a virtual photon $\gamma^\ast$ of momentum $q =l - l'$.  The DIS cross-section can be factorized into a convolution of the lepton tensor corresponding to the $\gamma^\ast$ emission by the electron and the hadron tensor describing the interaction of the virtual photon with the parton constituents of the hadron. The hadron tensor is the matrix element of the product of two electromagnetic currents $j^\mu = e_f \bar{\psi}\gamma^\mu \psi$ sandwiched between hadron states with momentum $P$ and spin $S$: 
\begin{eqnarray}
W^{\mu\nu}(q, P, S) = \frac{1}{2\pi} \int d^4 {\bf x} e^{iq\cdot {\bf x}} \langle P,S| j^\mu({\bf x}) j^\nu(0)|P,S\rangle\,.
\label{hT}
\end{eqnarray}
It can be expressed as the sum of its symmetric and antisymmetric parts,
\begin{eqnarray}
W^{\mu\nu}(q, P, S) = {\bar W}^{\mu\nu}(q, P) + i {\tilde W}^{\mu\nu}(q, P, S)\,,
\label{SAT}
\end{eqnarray}
where the symmetric part ${\bar W}$ does not depend on the hadron's spin unlike the antisymmetric part ${\tilde W}$. Each term in Eq. (\ref{SAT}) can be decomposed into all possible Lorentz structures, and after further considerations of gauge invariance, parity and time reversal invariance, can be expressed in terms of Lorentz invariant structure functions as 
 \begin{eqnarray}
\frac{1}{2}{\bar W}_{\mu\nu}(q, P) = \Big(-g_{\mu\nu} + \frac{q_\mu q_\nu}{q^2}\Big)F_1(x, Q^2) + \Big[\Big(P_\mu-\frac{P\cdot q}{q^2}q_\mu\Big)\Big(P_\nu-\frac{P\cdot q}{q^2}q_\nu\Big)\Big]\frac{F_2(x, Q^2)}{P\cdot q}\,,
\label{WS}
\end{eqnarray}
\begin{eqnarray}
\frac{1}{2}{\tilde W}_{\mu\nu}(q, P, S)= \frac{M}{P\cdot q}\epsilon_{\mu\nu\alpha\beta} q^\alpha\Big\{ S^\beta g_1(x, Q^2) + \Big[S^\beta - \frac{(S\cdot q)P^\beta}{P\cdot q}\Big]g_2(x, Q^2)\Big\}\,.
\label{WA}
\end{eqnarray}
We have used here the conventions $\epsilon^{0123} = -1$, $Q^2 = -q^2 > 0$, and introduced the Bjorken variable $x = Q^2/(2P\cdot q)$.

One can invert Eqs.~(\ref{WS}) and (\ref{WA}) and instead write down the structure functions in terms of the hadron tensor as
\begin{eqnarray}
F_1 = \Pi^{\mu\nu}_1 {\bar W}_{\mu\nu}\,,\ \ \  F_2 = \Pi^{\mu\nu}_2 {\bar W}_{\mu\nu}\,,
\label{F1-F2}
\end{eqnarray}
where 
\begin{eqnarray}
\Pi^{\mu\nu}_1=\frac{1}{4}\Big[\frac{1}{a}P^\mu P^\nu - g^{\mu\nu}\Big],
\ \ \ \Pi^{\mu\nu}_2=\frac{3P\cdot q}{4a}\Big[\frac{P^\mu P^\nu}{a}-\frac{1}{3}g^{\mu\nu}\Big]
\label{Proj}
\end{eqnarray}
are kinematic projectors, and $a=\frac{P\cdot q}{2x} + M^2\simeq \frac{P\cdot q}{2x}$. Similar relations for spin dependent structure functions can be found in Ref.~\cite{Leader:2001gr}.

In computing the hadron tensor, it is convenient to use its relation to the imaginary part of the forward Compton scattering amplitude,
\begin{eqnarray}
W^{\mu\nu} = \frac{1}{\pi}{\rm Im}\,T^{\mu\nu}\,,
\end{eqnarray}
where $T^{\mu\nu}$ is defined as the time ordered product
\begin{eqnarray}
T^{\mu\nu}(q, P, S) = i\int d^4 {\bf x}\, e^{iq\cdot {\bf x}}\langle P,S|T\{j^\mu({\bf x}) j^\nu(0)\}|P,S\rangle\,.
\label{Tprod}
\end{eqnarray}

Since QCD is intrinsically nonperturbative, one cannot compute the nonlocal matrix element in Eq. (\ref{Tprod}) directly. One can however make progress by writing this matrix element as a 
Taylor expansion when $x^\mu\rightarrow 0$, as the convolution of short distance perturbative coefficients $C_n(x^2,\mu)$ and long distance matrix elements of local operators $\langle\mathcal{O}_n^{\mu\nu}(0, \mu)\rangle$ separated at a factorization scale $\mu$; in this operator product expansion (OPE), 
\begin{eqnarray}
T\{j^\mu({\bf x}) j^\nu(0)\} \sim \sum_n C_n(x^2,\mu)\otimes \langle\mathcal{O}_n(0,\mu)^{\mu\nu}\rangle\,,
\label{OPE}
\end{eqnarray}
where $n$ here collectively denotes the spin of the local operator as well as the different operators having the same Lorentz structure.

The structure of Eq. (\ref{OPE}) can be understood from the point of view of the separation of kinematic modes with distinct time scales. Indeed a high energy scattering reaction is a combination of different subprocesses, each with its own spacetime scale. For example, in the OPE, the DIS interaction of the virtual photon with the target is determined by the large virtuality $Q^2$ which defines the size of the interaction area as $\sim 1/Q$. At the same time, the cross-section depends on the structure of the target as well, whose QCD dynamics is controlled by the nonperturbative scale $\Lambda_{\rm QCD}$. The coefficients $C_n$ contain the dynamics of the ``fast'' modes defined by the kinematics of the incoming photon and the matrix elements of local operators $\langle \mathcal{O}_n\rangle$ those of nonperturbative interactions inside the target hadron. Practical realizations of this factorization philosophy of the separation of kinematic modes can be far more involved and depend strongly on the process under consideration. 

In this paper, we wish to compute structure functions at small $x$ where the usual OPE of DIS breaks down~\cite{Mueller:1996hm}, but there nevertheless exists a strong separation of fast and slow modes in the rapidity variable. To understand this better, it is convenient to choose a frame for DIS where the target has zero transverse momentum and the longitudinal momentum component $P^+$ is large: $P = (P^+, M^2/2P^+, 0_\perp) \simeq (P^+, 0, 0_\perp)$. For the photon, we similarly set $q_\perp = 0$ with virtuality $Q^2 = -2 q^+q^- > 0$ and $x \simeq - q^+  / P^+ $. In the infinite momentum frame (IMF), $P^+\to \infty$ and $x<<1$ for fixed photon virtuality.  In the IMF, in analogy to the scale $\mu$ in the OPE, we can introduce a cutoff $\Lambda^+$ and define the ``slow'' modes  associated with the target as fields with $p^+ > \Lambda^+$, and likewise, the ``fast'' modes associated with the photon as fields with $p^+ < \Lambda^+$. 

\begin{figure}[htb]
 \begin{center}
 \includegraphics[width=120mm]{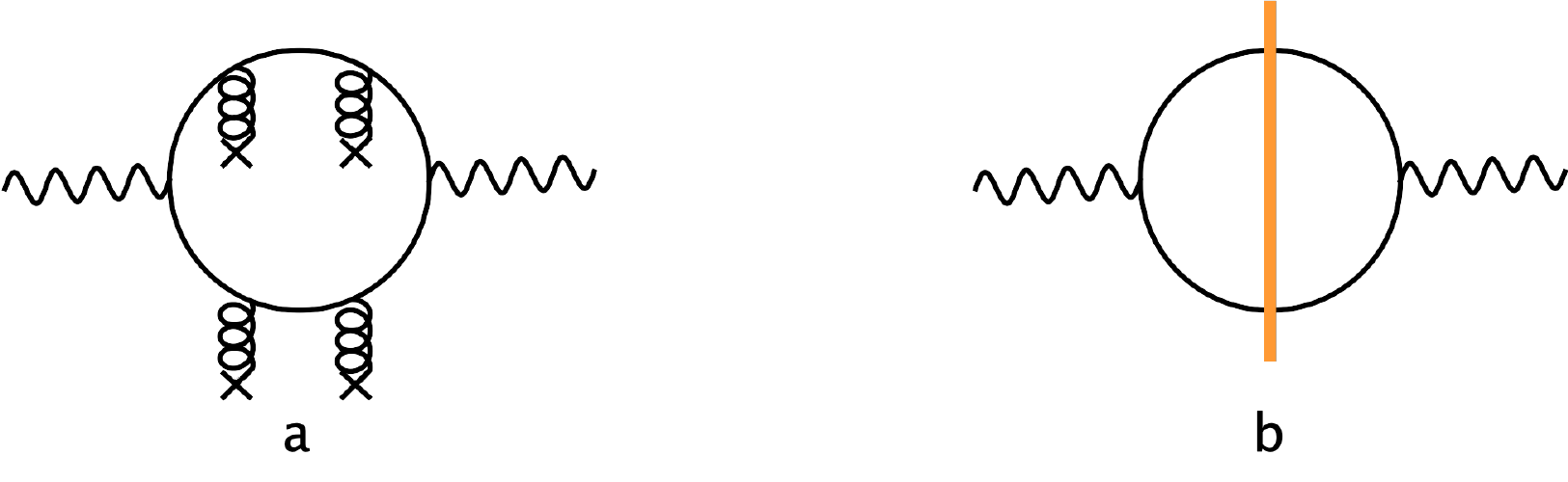}
 \end{center}
 \caption{\label{fig:1} a) Current-current correlator in an arbitrary background field; b) The same in the CGC shockwave background, where the spatial separation in $x^-$ shrinks to a point.}
 \end{figure}

In the Feynman diagram shown in  Fig. \ref{fig:1}a, the interaction of the fast modes with $p^+ < \Lambda^+$ is mediated by the photon virtuality $Q^2$ in the quark loop, and the integration over these fields can be computed explicitly using perturbation theory, yielding the equivalent of the coefficients $C_n$ in Eq. (\ref{OPE}). The vertical gluon lines, on the other hand, representing the slow  $p^+ > \Lambda^+$ modes, are absorbed into the operators $\mathcal{O}_n$ in Eq. (\ref{OPE}). In the following, we will refer to the gauge field corresponding to these modes as the background field.

At small $x$, the CGC EFT implements this separation of fast and slow modes. The high occupancy  background field gluon modes with transverse momenta $k_\perp \leq Q_s$ have high occupancy, and can therefore be obtained from solutions of the 
classical Yang-Mills equations~\cite{McLerran:1993ni, McLerran:1993ka, McLerran:1994vd,Kovchegov:1996ty},
\begin{eqnarray}
\mathcal{D}_\mu F^{\mu\nu} = J^\nu\,,
\label{EoM}
\end{eqnarray}
where the source $J^\mu$ describes the large $x$ modes in the ultrarelativistic hadron that in the IMF can be approximated as
\begin{eqnarray}
J^\nu = \delta^{\nu+}\delta(x^-)\rho(x^+,x_\perp)\,,
\end{eqnarray}
where $\rho(x^+,x_\perp)$ is the color charge density of the hadron. In covariant gauge, a static ($x^+$ independent) solution of the Yang-Mills equation exists, and is given by 
\begin{eqnarray}
A_{\rm cl}^+(x) = -\frac{1}{\partial^2_\perp} \rho(x_\perp)\delta(x^-);\ \ \ A_{\rm cl}^- = A_{\rm cl}^i = 0\,.
\label{BF}
\end{eqnarray}
This solution for the CGC shockwave background field has an infinitesimally small support of order $1/P^+$ in the $x^-$ direction, as represented in  Fig.~\ref{fig:1}b. 

It is clear from Fig. \ref{fig:1}b that there is a strong separation in time scales between the fast modes and the shockwave background field. As a result, the dominant contribution to the hadron tensor at small $x$ is from the diagram in Fig. \ref{fig:1}b, where the incoming photon splits into the quark-antiquark pair a long time before its interaction with the background~\cite{McLerran:1998nk}: 
\begin{eqnarray}
W^{\mu\nu}(q, P, S) = - \frac{e^2_f}{\pi}{\rm Im}\ i \int d^4x e^{iqx}\langle P,S| \gamma^\mu S_{\rm A_{\rm cl}}(x, 0)\gamma^\nu S_{\rm A_{\rm cl}}(0, x) \}|P,S\rangle\,,
\label{TprodSmx}
\end{eqnarray}
where $S_{\rm A_{\rm cl}}(x, 0)$ is the quark propagator in the CGC background field given by Eq.~(\ref{BF}). 

More generally, for an arbitrary background field $A$, defining the effective action $\Gamma$ as the functional integral
\begin{eqnarray}
e^{i\Gamma[A]} = \int \mathcal{D}\Psi \mathcal{D} \Psi^\dag e^{iS[\Psi, \Psi^\dag, A]} \,,
\label{EfAct}
\end{eqnarray}
one can express Eq.~(\ref{TprodSmx}) as\footnote{We will here understand the field $A$ as representing both the incoming photons and the non-Abelian gauge field background.} 
\begin{eqnarray}
W^{\mu\nu}(q, P, S) = \frac{1}{\pi e^2}{\rm Im}\  \int d^4x~ e^{iqx}\langle P,S| \frac{\delta^2 \Gamma[A]}{\delta A_\mu(x)\delta A_\nu(0)} |P,S\rangle\,,
\label{TprodEffex}
\end{eqnarray}
where the space-time metric has signature $g = (1, -1, -1, -1)$. The problem of computing structure functions is thereby reformulated as the problem of computing the effective action of the theory.
In this paper, we will develop a world-line framework for the computation of the effective action, and discuss results for structure functions at small $x$ for the case where the $A^\mu \rightarrow 
A_{\rm cl}^\mu$.  

\section{World-line framework for DIS}
\label{section:III}

We will present in this section an introduction to the world-line approach to initiate the unfamiliar reader to basic features of this formalism. We will begin with the case of scalar QED before proceeding to the full spinor QED case. We will then perform a perturbative computation of the vacuum polarization tensor; this computation will be useful for the computation in Section~\ref{section:IV}, where we 
will perform the computation of the same in the gluon shockwave background. Several details of the computations that may be of use to the interested reader are presented in Appendices~\ref{app:1}-\ref{app:5}. Some of the discussion in these introductory sections can be found in \cite{Schubert:2001he}. There are however a number of novel features that we have uncovered in our study that may of use in wider contexts. 

\subsection{World-line path integrals: Scalar QED}

The scalar QED  Lagrangian in a background field $A$ is 
\begin{eqnarray}
\mathcal{L}_{scalar} = \phi^\dag (\partial_\mu + ieA_\mu)^2 \phi - m^2 \phi^\dag \phi\,.
\end{eqnarray}

The effective action of the theory $\Gamma_{scalar}^M[A]$ can be defined as a functional integral over quantum fields $\phi$:
\begin{eqnarray}
e^{i\Gamma_{scalar}^M[A]} = \int \mathcal{D}\phi \mathcal{D} \phi^\dag e^{i\int d^4x \mathcal{L}_{scalar}}\,.
\label{MEfAction}
\end{eqnarray}

\begin{figure}[htb]
 \begin{center}
 \includegraphics[width=120mm]{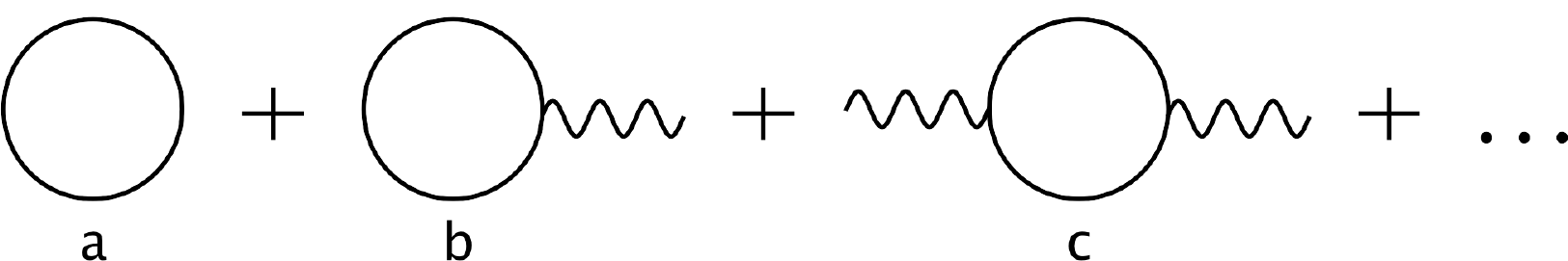}
 \end{center}
 \caption{The scalar QED effective action expanded in powers of the background field.\label{fig:2}}
 \end{figure}

To understand the structure of the effective action, one can construct the perturbative expansion of Eq. (\ref{MEfAction}), whereby the effective action is expressed as the sum of one particle irreducible diagrams (1PI). The first few terms of the perturbative expansion are represented in Fig. \ref{fig:2}. Since the dependence on quantum fields on the right hand side of Eq. (\ref{MEfAction}) has a quadratic form, the functional integrals can be easily evaluated;  the effective action of the theory then transforms into the functional determinant,
\begin{eqnarray}
i\Gamma_{scalar}^M[A] = \ln {\rm det} [- (\partial_\mu + i e A_\mu)^2 + m^2] = {\rm Tr}\ {\rm ln} [ - (\partial_\mu + i e A_\mu)^2 + m^2]\,,
\label{logdet}
\end{eqnarray}
where we use the well-known relation for the determinant of an arbitrary operator $\mathcal{O}$: $ \ln {\rm det} \mathcal{O} = {\rm Tr}\ {\rm ln} \mathcal{O}$. The trace in this relation should be understood as a discrete sum over finite dimensional  internal degrees of freedom (such as color and spin) and a functional trace over the continuous coordinate degrees of freedom. 

A standard way of computing  the functional determinant is based on the perturbative expansion of the logarithm on the right side of Eq. (\ref{logdet}), which corresponds to the resummation of the Feynman diagrams in Fig. \ref{fig:2}. The world-line approach presents an alternative definition of the functional determinant in the form of a one-dimensional functional integral. To construct such integral, one introduces a complete set of coherent states of the operator in Eq.~(\ref{logdet}) and writes the functional determinant as the product of the corresponding eigenvalues.

In order to do this, we first perform an analytical continuation of the effective action in Eq. (\ref{logdet}) to Euclidean space-time with $\eta^{\mu\nu} = {\rm diag}(1, 1, 1, 1)$ by Wick rotating the time variable $t_M\to -it_E$. That gives us the following form of the Euclidean effective action,
\begin{eqnarray}
\Gamma_{scalar}[A] = -{\rm Tr}\ {\rm ln} [(p_\mu + e A_\mu)^2 + m^2]\,.
\label{logdetE}
\end{eqnarray}
The operator in Eq. (\ref{logdetE}) has positive real eigenvalues, allowing us to apply the heat-kernel regularization formula:
 \begin{eqnarray}
 {\rm ln} [(p_\mu + e A_\mu)^2 + m^2] =  - \int^\infty_0 \frac{dT}{T} \Big(e^{-T [(p_\mu + e A_\mu)^2 + m^2] } - e^{-T }\Big).
 \label{hkernel}
\end{eqnarray}
The second term in this formula subtracts ultraviolet divergences; we will not need to consider these in the rest of this work. A general discussion of the regularization of UV divergences in the world-line formalism including, for instance, the derivation of the QCD $\beta$-function can be found in \cite{Schubert:2001he}. 
Substituting  Eq.~(\ref{hkernel}) into Eq. (\ref{logdetE}) gives,
\begin{eqnarray}
\Gamma_{scalar}[A] = {\rm Tr}\ \int^\infty_0 \frac{dT}{T} e^{-Tm^2} e^{-T (p_\mu + e A_\mu)^2 } .
\label{EA}
\end{eqnarray}
The exponential factor in this formula is understood as the evolution operator for a scalar particle world-line moving along a closed trajectory of length $T$.

We will now introduce the bosonic coherent states $|x\rangle$ and $|p\rangle$ that respectively define the state of a particle with coordinate $x$ and momentum $p$,  satisfying the following completeness and orthogonality relations:
\begin{eqnarray}
\int d^4x\, |x\rangle\langle x| = \int d^4p\, |p\rangle \langle p| = 1;\ \ \ \langle x|p\rangle = e^{ipx}.
\label{bosonort}
\end{eqnarray}
With these coherent states, the trace of an arbitrary operator $\mathcal{O}$ can be evaluated as
\begin{eqnarray}
{\rm Tr}\ \mathcal{O} = \int d^4x \langle x| \mathcal{O} |x \rangle\,,
\label{bosontrace}
\end{eqnarray}
which allows us to write the effective action in Eq.~(\ref{EA}) as 
\begin{eqnarray}
\Gamma_{scalar}[A] = \int^\infty_0 \frac{dT}{T} e^{-Tm^2} \int d^4x\, \langle x| e^{-T (p_\mu + e A_\mu)^2 }|x \rangle .
\label{EAafterTr}
\end{eqnarray}
This can equivalently be written as a functional integral by splitting the integral over $T$ into $N$ segments and inserting  complete sets of states $|x\rangle$ between them:
\begin{eqnarray}
\Gamma_{scalar}[A] = \int^\infty_0 \frac{dT}{T} e^{-Tm^2} \int_{PBC} \prod^N_{i=1} d^4x_i \langle x_{i+1}| e^{-\frac{T}{N} (p_\mu + e A_\mu)^2 }|x_i \rangle\,,
\label{EAprod}
\end{eqnarray}
where the integrals over $x_i$ satisfy periodic boundary conditions (PBC): $x_{N+1} = x_1$. Using the completeness and orthogonality relations in Eq.~(\ref{bosonort}), the matrix element of the evolution operator in Eq.~(\ref{EAprod}) can be rewritten as
\begin{eqnarray}
\langle x_{i+1}| e^{-\frac{T}{N} (p_\mu + e A_\mu)^2 }|x_i \rangle = \int \frac{d^4p_{i+1,i}}{(2\pi)^4} e^{i p_{i+1,i} (x_{i+1} - x_i )} \Big( 1 - (\tau_{i+1} - \tau_i) (p^{i+1,i}_\mu + e A^{i+1,i}_\mu )^2 + \dots \Big)\,,
\label{matrixbos}
\end{eqnarray}
where the ellipses stand for terms suppressed by higher powers of the ratio $T/N$. In Eq. (\ref{matrixbos}), we introduced a proper time variable $\tau$ such that $\tau_1 = 0$, $\tau_{N+1} = T$ and $\tau_{i+1} - \tau_i = T/N$. The evolution operator in Eq.~(\ref{matrixbos}) depends on the background field, whose magnitude is evaluated the average value  $\bar{x}_{i+1, i} = (x_{i+1} - x_i)/2$, namely, $A^{i+1,i}_\mu = A_\mu(\bar{x}_{i+1,i})$.

Substituting Eq.~(\ref{matrixbos}) into Eq.~(\ref{EAprod}), and taking the limit $N\to \infty$, yields finally the functional integral representation of the effective action to be 
\begin{eqnarray}
\Gamma_{scalar}[A] = \int^\infty_0 \frac{dT}{T} e^{-Tm^2} \int_{PBC} \mathcal{D}x \int \mathcal{D}p \ \mathcal{P} \exp\Big\{ \int^T_0 d\tau \Big(i p \dot{x} - (p_\mu + e A_\mu )^2\Big) \Big\}\,.
\label{EAfunc}
\end{eqnarray}
Note that now $x(\tau)$ is a function of the proper time $\tau$ and that the functional integral $\int \mathcal{D} x$ is performed over all possible configurations of $x(\tau)$. The operator $\mathcal{P}$ imposes path ordering of the background fields $A$ along $x(\tau)$. For the sake of brevity, we will not write it in the equations that follow but path ordering will always be implicit in our discussion.

Observe that the effective action in Eq.~(\ref{EAfunc}) has a structure distinct from that in the standard quantum field theory expression given in Eq.~(\ref{MEfAction}). Instead of quantum fields,  currents, and products of currents (obtained from taking functional derivations of the effective action with respect to the external fields), are expressed in terms of an embedded $0+1$-dimensional quantum mechanical probe. The 0+1-dimensional quantum mechanical world-line trajectory $x(\tau)$ is the novel ingredient in this approach that makes it distinct from the usual description in terms of quantum fields. As we will see later, this allows one to develop effective techniques to calculate world-line functional integrals employing quantum mechanical world-line propagators.

We note finally that since the functional integral over momentum in Eq.~(\ref{EAfunc}) is Gaussian, it can be evaluated easily after performing a proper shift of variables to give,
\begin{eqnarray}
\Gamma_{scalar}[A] = \int^\infty_0 \frac{dT}{T} e^{-m^2T} \int_{PBC} \mathcal{D}x \ \exp\Big\{ - \int^T_0 d\tau \mathcal{L}_{scalar} \Big\}\,,
\label{EAfuncFin}
\end{eqnarray}
with the world-line Lagrangian defined as 
\begin{eqnarray}
\mathcal{L}_{scalar} = \frac{1}{4}\dot{x}^2 + ie\dot{x}\cdot A \,. 
\end{eqnarray}
The first term in this expression is the kinetic term for a  free world-line. The second coupling term represents the evolution of the world-line in the background field $A$. As has been noted previously, this term is equivalent to a Wilson loop of the background field $A$ -- and therefore is gauge invariant with respect to the background field~\cite{Barducci:1976xq,Balachandran:1976ya,Barducci:1980xk,Strassler:1992zr}.

\subsection{World-line path integrals: Spinor QED}

One can similarly construct the world-line representation of the effective action for spinor QED. The additional novel ingredient that deserves discussion is the introduction of the fermion coherent 
states $|\xi\rangle$ and $|\bar{\xi}\rangle$ to describe spinor fields in the effective action. The interested reader can find the details of the calculation in Appendix \ref{app:1}. The final expression for the effective action of spinor QED derived there is 
\begin{eqnarray}
\Gamma_{QED}[A] = -\frac{1}{2} \int^\infty_0 \frac{dT}{T} e^{-m^2T} \int_{PBC} \mathcal{D}x \int_{APBC} \mathcal{D} \psi \exp\Big\{-\int^T_0 d\tau \Big(\frac{1}{4} \dot{x}^2 + \frac{1}{2}\psi_\mu\dot{\psi}^\mu + ie\dot{x}^\mu A_\mu - ie \psi^\mu \psi^\nu F_{\mu\nu}\Big)\Big\}\,.
\label{EAQEDWl}
\end{eqnarray}

In analogy to scalar QED, the world-line trajectory of a particle is described by both bosonic coordinate $x(\tau)$ and fermionic Grassmann $\psi(\tau)$ point particle degrees of freedom. As discussed at length in Appendix \ref{app:1}, the Grassmann world-line trajectory $\psi(\tau)$ is obtained from the product of eigenvalues of the QED kinetic operator in the Hilbert space defined by the  fermionic coherent states $|\xi\rangle$. While the functional integral over $x$ has periodic boundary conditions (PBC) $x(0) = x(T)$, the Grassmann functional integral instead has  
antiperiodic boundary conditions (APBC) $\psi(0) = -\psi(T)$. The world-line interactions with the background field are now given by the coupling terms $\dot{x}^\mu A_\mu$ and $\psi^\mu \psi^\nu F_{\mu\nu}$. The first term coincides with the spin-independent term in Eq. (\ref{EAfuncFin}) while the second term is unique for the case of spinor QED and describes the transmission of polarization from the background field to the world-line. This point will be discussed further in \cite{inprep:2019}. Note that the bosonic and fermionic sectors of the theory are not independent due to the dependence of the strength tensor $F_{\mu\nu}$ on $x(\tau)$.

However the two sectors are independent in the case of a free world-line, as represented by the  kinetic terms in Eq. (\ref{EAQEDWl}). Introducing the notation,
\begin{eqnarray}
G^{-1}_B(\tau_1, \tau_2) = \frac{1}{2}\frac{\partial^2}{\partial\tau^2_1}\delta(\tau_1 - \tau_2)\,\,;\ \ \ G^{-1}_F(\tau_1, \tau_2) = \frac{1}{2}\frac{\partial}{\partial\tau_1}\delta(\tau_1 - \tau_2)\,,
\label{props}
\end{eqnarray}
the kinetic terms in the effective action (\ref{EAQEDWl}) (corresponding to the diagram in Fig. \ref{fig:2}a) can be rewritten as 
\begin{eqnarray}
&&\Gamma_{QED}[A = 0] = -\frac{1}{2} \int^\infty_0 \frac{dT}{T} e^{-m^2T} \int_{PBC} \mathcal{D}x \exp\Big\{ \frac{1}{2} \int^T_0 d\tau d\tau' x^\mu(\tau)G^{-1}_B(\tau, \tau')x_\mu(\tau') \Big\} 
\nonumber\\
&&\times \int_{APBC} \mathcal{D} \psi \exp\Big\{-\int^T_0 d\tau d\tau' \psi^\mu(\tau) G^{-1}_F(\tau, \tau' )\psi_\mu(\tau') \Big\}
\label{EAQEDWlfree}
\end{eqnarray}
The functions $G_B(\tau_1, \tau_2)$ and $G_F(\tau_1, \tau_2)$ are respectively bosonic and fermionic propagators of the world-line satisfying the corresponding periodic and anti-periodic boundary conditions on a circle of circumference $T$. The explicit forms of their solutions~\cite{Bern:1991aq,Strassler:1992zr} are 
\begin{eqnarray}
G_B(\tau, \tau') = |\tau - \tau'| - \frac{(\tau - \tau')^2}{T};\ \ \ G_F(\tau, \tau') = {\rm sign}(\tau - \tau')\,,
\label{wlprop}
\end{eqnarray}
and that of their derivatives are 
\begin{eqnarray}
\partial_\tau G_B(\tau, \tau') = {\rm sign}(\tau - \tau') - \frac{2(\tau - \tau')}{T};\ \ \ \partial_\tau G_F(\tau, \tau') = \delta(\tau - \tau')\,,
\label{wlpropDer}
\end{eqnarray}
and
\begin{eqnarray}
\partial^2_\tau G_B(\tau, \tau') = 2\delta(\tau - \tau') - \frac{2}{T}\,.
\label{wlpropDer2}
\end{eqnarray}

Since the world-line functional integrals in Eq. (\ref{EAQEDWlfree}) are quadratic, they can be evaluated explicitly and give  
\begin{eqnarray}
\int \mathcal{D} x \exp\Big\{- \int^T_0 d\tau \frac{1}{4} \dot{x}^2\Big\} = (4\pi T)^{-D/2}\,;\ \ \ \int \mathcal{D}\psi \exp\Big\{-\int^T_0 d\tau \frac{1}{2}\psi\dot{\psi}\Big\} = 4 \,,
\label{freefunc}
\end{eqnarray}
where $D$ represents the number of spacetime dimensions.

Calculating functional integrals in the general case of the interacting world-line effective action in Eq.~(\ref{EAQEDWl}) is a formidable task because of the arbitrary dependence of the background field $A$ on the coordinate. However one can use the formulas in Eq.~(\ref{freefunc}) and (\ref{wlprop}) to construct a perturbative expansion of the effective action. In the next section, we will give the simplest example of such a calculation, the computation of the photon polarization tensor in QED.  We will later generalize this computation to the more complicated case of the interaction of the world-lines with the small $x$ gluon shockwave background.

The generalization of Eq. (\ref{EAQEDWl}) to the case of QCD is straightforward: one should promote the photon fields $A$ to include as well fields with color degree of freedom and add the trace over color ${\rm Tr}_c$ in front of the whole expression. One can also use fermionic coherent states to construct the functional integral representation for the trace over color~ \cite{Barducci:1980xk,DHoker:1995uyv,DHoker:1995aat}. We have considered colored Grassmann point like variables recently in the context of kinetic theory~\cite{Mueller:2019gjj}; we will however not employ them here but will return to them in future.  One should also note that the ordering of matrices in the color trace is fixed by the proper time ordering of the corresponding gluon fields.

\subsection{Vacuum polarization tensor in the world-line approach}
We will now consider as a simple example of a perturbative calculation in the world-line formalism, the computation of the photon polarization tensor in the vacuum. We will sketch the main ingredients of the calculation; these will be useful later for the more complicated case of the quark loop in the gluon shock-wave background.

In general, to take into account the interaction of the world-line with the background field $A$, one should be able to calculate the functional integrals over the world-line trajectories in Eq.~(\ref{EAQEDWl}). Unfortunately, since this cannot be performed exactly for an arbitrary background field, one needs to develop approximate computational schemes. One approach is 
to employ a semi-classical approximation whereby the world-line functional integrals are expanded around classical trajectories defined by the world-line Euler-Lagrange equations of motion:
\begin{eqnarray}
\bar{m}\ddot{x }^{cl}_\mu = ig F_{\mu\nu}\cdot \dot{x }^{cl}_\nu - \frac{ig}{2\bar{m}} \psi^{cl}_\rho \partial_\mu F_{\rho\sigma} \psi^{cl}_\sigma\,;\ \ \ \bar{m} \dot{\psi}^{cl}_\mu = igF_{\mu\sigma} \psi^{cl}_\sigma\,,
\label{eqmfull}
\end{eqnarray}
where $\bar{m}^2 \equiv m^2 + i\int^1_0 du \psi^\mu F_{\mu\nu}\psi^\nu $. These equations are the covariant generalizations of the Bargmann-Michel-Telegdi (BMT) equations for spinning particles in external gauge fields \cite{Bargmann:1959gz}. 

In this paper, for the problem at hand,  we will use an approach based on the perturbative expansion of the exponent in Eq.~(\ref{EAQEDWl}). This method \cite{Strassler:1992zr} is an alternative perturbative approach to the computation of the standard Feynman diagrams in quantum field theory and leads to the same results albeit presented in a quite different form. 
A powerful feature of the approach is the efficient computation of spinor and color traces. Despite this, the world-line approach is not widely applied to address problems in QCD. 
Our work takes a step towards redressing this situation by performing novel practical computations. 
 
We begin our discussion of vacuum polarization by considering the simple diagram with two background photons given in Fig. \ref{fig:2}c. We first take the functional derivative of Eq.~(\ref{EAQEDWl}) and take its Fourier transform: 
\begin{eqnarray}
&&\int d^4z_1 \frac{\delta \Gamma[A]}{\delta A_\mu(z_1)} e^{ik_1z_1} \Big|_{A=0} = -\frac{ ie}{2}\int^\infty_0\frac{dT}{T} e^{-m^2T}\int \mathcal{D}x\int \mathcal{D}\psi \int^T_0 d\tau_1 (\dot{x}^\mu_1 + 2i\psi^\mu_1\psi^\rho_1 k_{1\rho}) e^{ik_1 x_1}\ e^{-\int^T_0 d\tau(\frac{1}{4}\dot{x}^2 + \frac{1}{2}\psi\dot{\psi})}
\label{curfunc}
\end{eqnarray}
where we introduced the short-hand notation $x_1 \equiv x(\tau_1)$, $\psi_1 \equiv \psi(\tau_1)$. In Eq. (\ref{curfunc}), the world-line current,
\begin{eqnarray}
j^\mu = \dot{x}^\mu + 2i\psi^\mu \psi^\rho k_\rho\,,
\label{current}
\end{eqnarray}
describes the interaction of the world-line with an incoming photon carrying momentum $k$. Since the interaction can occur at any value of the proper time $\tau_1$, we have integrated it over 
in  Eq. (\ref{curfunc}). In the convolution of Eq. (\ref{curfunc}) with momentum $k_\mu$,  the second term in the world-line current vanishes due to the Grassmann nature of the trajectory $\psi(\tau)$. Further, it is also easy to show that the convolution of the $\dot{x}$ term with $k_\mu$ leads to a full derivative over time $\tau_1$, which in turn is trivially zero because the functional integral over the coordinate $x$ satisfies periodic boundary conditions. This world-line current is therefore manifestly conserved. Note that one can  interpret the first term of Eq. (\ref{current}) as the {\it scalar} current corresponding to the interaction of the scalar world-line with the background field and the second term as its {\it spinor} counterpart.

The photon polarization diagram in Fig. \ref{fig:2}c corresponds to the Fourier transform of the second derivative of the effective action, 
\begin{eqnarray}
\Gamma^{\mu\nu}[k_1, k_2] = \int d^4z_1 d^4z_2 \frac{\delta^2 \Gamma[A]}{\delta A_\mu(z_1)\delta A_\nu(z_2)}|_{A=0} \,e^{ik_1 z_1} e^{ik_2 z_2}\,,
\label{secder}
\end{eqnarray}
which can be written as 
\begin{eqnarray}
&&\Gamma^{\mu\nu}[k_1, k_2] = -\frac{(ie)^2}{2}\int^\infty_0\frac{dT}{T} e^{-m^2T}\int \mathcal{D}x\int \mathcal{D}\psi \prod^2_{i=1} \int^T_0 d\tau_i (\dot{x}^{\eta_i}_i + 2i\psi^{\eta_i}_i \psi^\rho_i k_{i\rho}) e^{ik_i x_i}
e^{-\int^T_0 d\tau(\frac{1}{4}\dot{x}^2 + \frac{1}{2}\psi\dot{\psi})}\,,
\label{polarization}
\end{eqnarray}
where $k_1$ and $k_2$ are two incoming momenta, and the Lorentz indices $\eta_1 = \mu$, and $\eta_2 = \nu$. The problem of computing the vacuum polarization tensor $\Gamma^{\mu\nu}[k_1, k_2]$ is therefore equivalent to computing the functional integrals in the expression above.

A discussion of the computation of world-line functional integrals is provided in Appendixes \ref{app:2} and \ref{app:3}. Taking into account that
\begin{eqnarray}
&&\int \mathcal{D}\psi \,\psi^\mu_1  \psi^\rho_1e^{-\int^T_0 d\tau (\frac{1}{2}\psi \dot{\psi})} = 0\,,
\end{eqnarray}
we find that there are only two nontrivial terms:
\begin{eqnarray}
&&\int \mathcal{D}x\int \mathcal{D}\psi (\dot{x}^\mu_1 + 2i\psi^\mu_1 \psi^\rho_1 k_{1\rho})(\dot{x}^\nu_2 + 2i\psi^\nu_2 \psi^\sigma_2 k_{2\sigma}) e^{ik_1 x_1}e^{ik_2 x_2}
e^{-\int^T_0 d\tau(\frac{1}{4}\dot{x}^2 + \frac{1}{2}\psi\dot{\psi})}
\nonumber\\
&&= \int \mathcal{D}x\int \mathcal{D}\psi \sum^2_{i=1} \Delta^{\mu\nu}_i e^{ik_1 x_1}e^{ik_2 x_2}
e^{-\int^T_0 d\tau(\frac{1}{4}\dot{x}^2 + \frac{1}{2}\psi\dot{\psi})}\,,
\label{PolarTerms}
\end{eqnarray}
where
\begin{eqnarray}
\Delta^{\mu\nu}_1 = \dot{x}^\mu_1\dot{x}^\nu_2\,\,\,;\ \ \ \Delta^{\mu\nu}_2 = (2i\psi^\mu_1 \psi^\rho_1 k_{1\rho} )(2i\psi^\nu_2 \psi^\sigma_2 k_{2\sigma} )\,.
\label{Delt1}
\end{eqnarray}

One thereby obtains the compact expression,
\begin{eqnarray}
\Gamma^{\mu\nu}[k_1, k_2] = (2\pi)^4\delta^4(k_1 + k_2) \sum^2_{i=1}\Pi^{\mu\nu}_i(q)\,,
\end{eqnarray}
where $q\equiv k_1$ and the polarization tensors $\Pi^{\mu\nu}_1$ and $\Pi^{\mu\nu}_2$ respectively correspond to the terms $\Delta^{\mu\nu}_1$ and $\Delta^{\mu\nu}_2$ in Eq. (\ref{Delt1}).

In Appendix \ref{app:2}, we provide a detailed derivation of the corresponding scalar world-line functional integrals. Employing these results, we obtain  
\begin{eqnarray}
&&\Pi^{\mu\nu}_1(q) = 2 (ie)^2 \int\frac{d^4p}{(2\pi)^4} \frac{-\eta^{\mu\nu}\{(p + q)^2 + m^2\} - \eta^{\mu\nu}\{p^2 + m^2\} + (2p^\mu + q^\mu)(2p^\nu + q^\nu)}{\{p^2 + m^2\}\{(p + q)^2 + m^2\}}\,.
\label{P1Vac}
\end{eqnarray}
Likewise, in Appendix \ref{app:3}, we derive explicit expressions for the Grassmann world-line functional integrals, leading to the result 
\begin{eqnarray}
&&\Pi^{\mu\nu}_2[q] = 2(ie)^2 \int\frac{d^4p}{(2\pi)^4} \frac{\eta^{\mu\nu} q^2 - q^\mu q^\nu}{\{p^2 + m^2\}\{(p+q)^2 + m^2\}}\, .
\label{P2Vac}
\end{eqnarray}
The connection of results from the world-line computation with standard Feynman diagram techniques is discussed at length in Appendix \ref{app:5}. In particular, the first two terms in Eq.~(\ref{P1Vac}) correspond to delta functions in the second derivative of the world-line propagator given in Eq. (\ref{wlpropDer2}).

Finally, we take the sum of Eqs. (\ref{P1Vac}) and (\ref{P2Vac}) and analytically continue to Minkowski space with signature $g = (1, -1, -1, -1)$. This is achieved by the replacements
\begin{eqnarray}
\eta_{\mu\nu} \to -g_{\mu\nu},\ \ \ \ \ \ k^4 \to -ik^0\,.
\label{acont}
\end{eqnarray}
With these replacements, we obtain 
\begin{eqnarray}
&&i\Pi^{\mu\nu}[q] = -4e^2 \int\frac{d^4p}{(2\pi)^4} \frac{p^\mu (p^\nu + q^\nu) + (p^\mu + q^\mu) p^\nu - g^{\mu\nu}\{p^2 + (p\cdot q) - m^2\}}{\{p^2 - m^2 + i\epsilon\}\{(p + q)^2 - m^2 + i\epsilon\}}\,.
\label{vacRes}
\end{eqnarray}
The numerator of this expression is easily identified as the trace ${\rm Tr}\{\gamma^\mu (\slashed{p} + m)\gamma^\nu (\slashed{p} + \slashed{q} + m)\}$ in the Feynman diagram calculation; our result therefore coincides with the standard expression for the photon polarization tensor.
We see therefore that the world-line formalism is an efficient way to compute the effective action and its functional derivatives. Its advantages are not manifest for simple spinor traces; however 
because of the relative ease in performing quantum mechanical Grassmann integrals, these advantages become more apparent in the computation of more complex traces.  

In the next section, we will generalize the calculation presented here of the polarization tensor in the vacuum to the case of interaction  of the quark loop with the gluon shockwave background field; we will use this to calculate the hadron tensor $W^{\mu\nu}$ in Eq. (\ref{TprodEffex}). As we shall see, the result for the polarization tensor will be similar to Eq. (\ref{polarization}), but will now include world-line currents corresponding to the interaction of the world-lines with the shockwave background. Further, integration over world-line trajectories will lead to formulas similar to  Eqs. (\ref{P1Vac}) and (\ref{P2Vac}), albeit the expressions are more involved in that case.

\section{World-lines and the shockwave approximation}
\label{section:IV}
The propagation of the world-line in the background field is described in coordinate space. This makes the approach particularly suitable for calculations in the shockwave approximation because in this approximation the interaction is localized at $x^-=0$. In this section, we will show how the shockwave approximation can be implemented in the world-line approach and we shall introduce the building blocks which define the world-line trajectory in the CGC background. In the next section, we will use it to compute the spin averaged structure functions $F_1$ and $F_2$ that are the Lorentz invariant quantities in the general formula for the hadron tensor in Eq.~(\ref{TprodEffex}).

We begin our discussion with the world-line effective action $\Gamma[A]$ for QCD which can be easily obtained\footnote{Note that we have omitted here additional terms in the effective action representing the coupling of quarks to the external electromagnetic field.} from Eq. (\ref{EAQEDWl}) by insertion of the color trace over color indexes of the background fields:
\begin{eqnarray}
&&\Gamma_{QCD}[A] = -\frac{1}{2} \int^\infty_0 \frac{dT}{T} e^{-m^2T}\, {\rm Tr_c} \int \mathcal{D}x \int \mathcal{D} \psi \exp\Big\{-\int^T_0 d\tau \Big(\frac{1}{4} \dot{x}^2 + \frac{1}{2}\psi_\mu\dot{\psi}^\mu + ig\dot{x}^\mu A_\mu - ig \psi^\mu \psi^\nu F_{\mu\nu}\Big)\Big\}\,,
\label{MLag}
\end{eqnarray}
where we assume periodic and anti-periodic boundary conditions for the functional integrals over $x(\tau)$ and $\psi(\tau)$ coordinates respectively.

From this formula, it is easy to obtain the second derivative of the effective action (with respect to the electromagnetic field) which defines the structure of the hadron tensor in the kinematic limit of small $x$; it is defined by the product of the  two world-line currents in Eq.~(\ref{current}) and has the form
\begin{eqnarray}
&&\Gamma^{\mu\nu}[k_1, k_3] = \frac{e^2 e_f^2 }{2}\int^\infty_0\frac{dT}{T} e^{-m^2T}~{\rm Tr_c} \int \mathcal{D}x\int \mathcal{D}\psi \int^T_0 d\tau_1 \int^T_0 d\tau_3
\nonumber\\
&&\times~ (\dot{x}^\mu_1 + 2i\psi^\mu_1 \psi^\rho_1 k_{1\rho}) \,e^{ik_1 x_1}~(\dot{x}^\nu_3 + 2i\psi^\nu_3 \psi^\sigma_3 k_{3\sigma}) \,e^{ik_3 x_3}\exp\Big\{-\int^T_0 d\tau \Big(\frac{1}{4} \dot{x}^2 + \frac{1}{2}\psi_\mu\dot{\psi}^\mu + ig\dot{x}^\mu A_\mu - ig \psi^\mu \psi^\nu F_{\mu\nu}\Big)\Big\}.
\label{photamp}
\end{eqnarray}
As in the QED case, the formula describes the interaction of the world-line with two incoming photons with momenta $k_1$ and $k_3$, that respectively interact with the world-line trajectory with the charge $e_f$ (in units of the electromagnetic charge) of a given quark flavor at the proper times $\tau = \tau_1$ and $\tau = \tau_3$. (As usual, the subscript on the variables denotes their values at a particular $\tau$, $x_1 \equiv x(\tau_1)$, etc.)

Here we wish to take into account the interactions of the world-line with the CGC background field $A^\mu(x)$, where the interaction terms are specified by $\dot{x}^\mu A_\mu(x)$ and $\psi^\mu \psi^\nu F_{\mu\nu}(x)$ in the world-line action.
The CGC background field is given by the solution to the Yang-Mills equations in Eq. (\ref{BF}) which has a shock-wave structure at high energies. Namely, a segment of the world-line interacts with the CGC background instantaneously and multiple interactions with the background gluons shrink to a single point. From Fig. \ref{fig:1}b, it is clear that there are only two such points on the world-line which we denote by the proper-time variables $\tau_2$ and $\tau_4$. We will now consider {\it only one} of these interaction points and show how the instantaneous nature of the CGC background modifies the two interaction terms in the world line action in Eq.~(\ref{photamp}) that represents the interaction with the background field.

Let us start with the Grassmann phase factor. The instantaneous nature of the world-line interaction with the CGC background dramatically simplifies this expression. To illustrate this, we can expand the phase factor in powers of $\psi$'s,
\begin{eqnarray}
&&e^{ig\int^T_0 d\tau \psi \psi F} = \sum^\infty_{N=0} \frac{(ig)^N}{N!} \prod^N_{n=1}\int^T_0 d\tau_n \psi^\mu_n \psi^\nu_n F_{\mu\nu}(x_n)\,.
\label{GrasmExp}
\end{eqnarray}
It is easy to see that only first two orders of the expansion are important. Indeed, from Eq. (\ref{BF}), we find that the interaction with the CGC background shrinks to a point at $x^- = 0$, which corresponds to a local interaction in the world-line trajectory at a proper time $\tau_i$. As a result, the factors $\psi^\mu_n\psi^\nu_n$ in Eq. (\ref{GrasmExp}) are taken at the same point of the world-line ($\tau_n = \tau_i$). Thus due to the Grassmann nature of the variables, only the first three terms in the expansion survive.

According to Eq. (\ref{BF}) the shockwave background field has only one non-zero component $A_-(x)$. However in general, at high energies, we can assume that there is also a non-trivial sub-eikonal component $A_\perp(x)$~\cite{Balitsky:2017flc}. 
In the Regge limit of QCD, this $A_i$ component describes the transition of polarization from gauge fields to fermion degrees of freedom. As a result, our derivation below in principle includes not only the scattering in the CGC background (\ref{BF}) but also describes spin effects at small $x$.

This discussion indicates that the field strength tensor of the background field has only two non-zero components $F_{-m}$ and $F_{mn}$. Therefore the terms of expansion in Eq. (\ref{GrasmExp}) with $N>1$ at a given time $\tau_i$ (which is $\tau_2$ or $\tau_4$ in our notations) can only generate trivial structures, for example $\psi^-_i\psi^m_i \psi^n_i \psi^l_i = 0$, where $m$, $n$ and $l$ are transverse Lorentz indices. 
As a result for a single interaction of the world-line with the CGC background (which of course includes infinite number of background gluons) the Grassmann phase factor in Eq.~(\ref{GrasmExp}) simplifies to
\begin{eqnarray}
1 + ig \int^T_0 d\tau_i \psi^\eta_i \psi^\kappa_i F_{\eta\kappa}(x_i) \,.
\label{oneintCGC}
\end{eqnarray}

\begin{figure}[htb]
 \begin{center}
 \includegraphics[width=100mm]{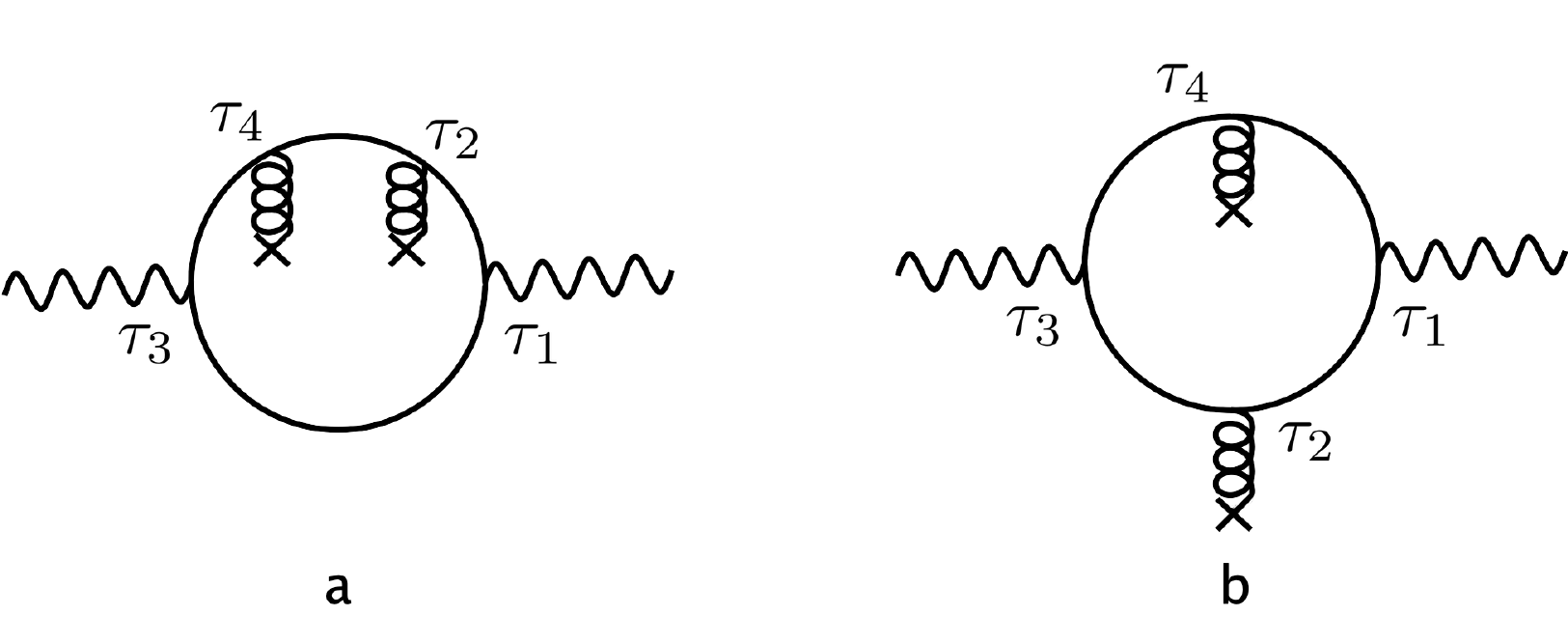}
 \end{center}
 \caption{\label{fig:3}Functional integrals with different time orderings. Each vertical gluon line denotes  the interaction with an infinite number of background gluons that are shrunk to a single point on the world-line.}
 \end{figure}

Now we need to take into account  the fact that the world-line trajectory can interact with the CGC background more than once. Both segments, between points $\tau_1$ and $\tau_3$, interact with the external field, which is obvious from Fig. \ref{fig:1}b. The Grassmann phase factor of each of these interactions is described by Eq.~(\ref{oneintCGC}). 
We can therefore rewrite the second derivative of the effective action Eq. (\ref{photamp}) as\footnote{As usual, we will assume path ordering of fields along the world-line trajectory--see the discussion after Eq. (\ref{EAfunc}).}
\begin{eqnarray}
&&\Gamma^{\mu\nu}[k_1, k_3] = \frac{e^2 e_f^2 }{2}\int^\infty_0\frac{dT}{T} e^{-m^2T}~{\rm Tr_c} \int \mathcal{D}x\int \mathcal{D}\psi \int^T_0 d\tau_1 \int^T_0 d\tau_3~(\dot{x}^\mu_1 + 2i\psi^\mu_1 \psi^\rho_1 k_{1\rho})~e^{ik_1 x_1}~(\dot{x}^\nu_3 + 2i\psi^\nu_3 \psi^\sigma_3 k_{3\sigma})~e^{ik_3 x_3}
\nonumber\\
&&\times ~\Big(1 + i g \int^T_0 d\tau_2    \psi^\xi_2 \psi^\lambda_2 F_{\xi\lambda}(x_2) \Big) \Big(1 + i g \int^T_0 d\tau_4 \psi^\eta_4 \psi^\kappa_4 F_{\eta\kappa}(x_4)\Big)~ \exp\Big\{-\int^T_0 d\tau \Big(\frac{1}{4} \dot{x}^2 + \frac{1}{2}\psi_\mu\dot{\psi}^\mu + ig\dot{x}^\mu A_\mu \Big)\Big\},
\label{masterWF}
\end{eqnarray}
where $\tau_2$ and $\tau_4$ are the proper times corresponding to the interaction of the world-line trajectory with the CGC shockwave. This is shown in Fig. \ref{fig:3} where the  vertical gluon line drawn denotes the interaction of the world-line with an infinite number of background gluons that are shrunk to a single point on the world-line-as also shown in the representation of Fig. \ref{fig:1}b.

In the above expression, we integrated over all possible values of $\tau_2$ and $\tau_4$, from $0$ to $T$. A  configuration for the polarization tensor in shockwave background can for instance correspond to the world-line interacting with the virtual photon at $\tau_1$, then with the shockwave at $\tau_2$ with $x^-_2 = 0$; it can then fly off to $x^-\to\infty$ as it were a free particle, return from $x^-=\infty$, interact with the shockwave again at $\tau_4$, and finally interact again with the virtual photon at $\tau_3$, as depicted in Fig. \ref{fig:3}a. 

\begin{figure}[htb]
 \begin{center}
 \includegraphics[width=170mm]{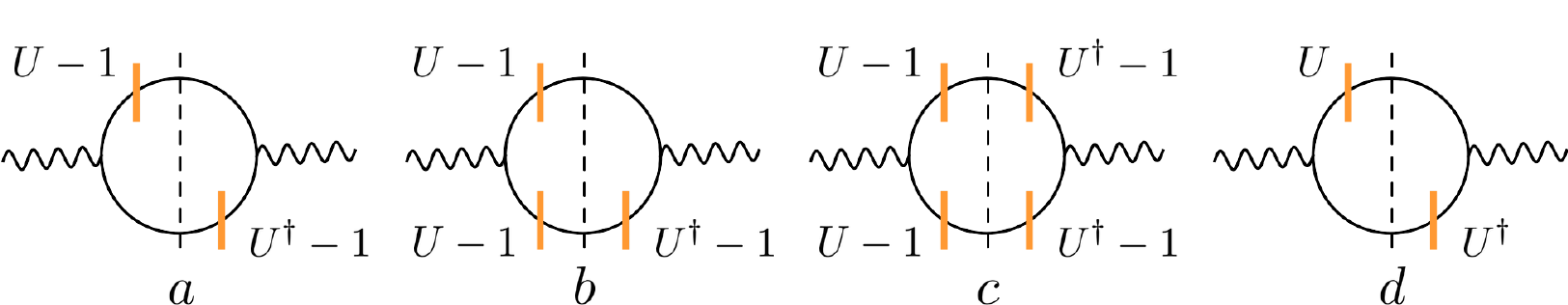}
 \end{center}
 \caption{a-c) The structure of the phase factors acquired by the world-line due to the interaction of the $q\bar{q}$ pair with the shockwave background field. The vertical dashed line is the cross-section cut, which is defined by the world-line going to $x^- \to \infty$.  d) The effective sum of these contributions. \label{fig:7}}
 \end{figure}

In principle, more involved configurations are possible, where the world-line trajectory can intersect with the CGC background up to four times. However one can show that due to cancellations between different phase factors, only two interactions survive. This is particularly easy to see in the case of scalar QED where the phase factors have a simple form $U(x_\perp)\equiv U_{[\infty, -\infty]}(x_\perp)$ defined in Eq. (\ref{phasefactors}). In Figs. \ref{fig:7}a-c we present typical combinations of the phase factors acquired by the world-line in the DIS cross section. Each factor $U-1$ describes multiple interactions of the world-line with the shockwave background. Note that in the DIS when $q^2 <0$ there should be at least one interaction with the background on  each side of the cut. Using unitarity of the phase factors, $UU^\dag = 1$, one finds that the resummation of contributions in Figs. \ref{fig:7}a-c leads to the structure of factors shown in Fig. \ref{fig:7}d. The latter can be interpreted as two interactions of the world-line with the shockwave background defined by the phase factor $U$. We will introduce this factor later in the world-line currents  in Eqs. (\ref{efcurwithph}) and (\ref{efcurwithph2}).  

By definition (see Eq. (\ref{BF})), the points of the world-line $\tau_2$ and $\tau_4$ are located at $x^- = 0$. In our convention, these are represented as $x^-_{2,4}=0$.  It is convenient to make this explicit by using the identity,
\begin{eqnarray}
&&\int^T_0 d\tau_{2,4}~ \dot{x}^-_{2,4}\, {\rm sign}(\dot{x}^-_{2,4}) \,\delta(x^-_{2,4}) = 1\,,
\end{eqnarray}
and rewriting the amplitude in Eq.~(\ref{masterWF}) as\footnote{Strictly speaking, the effective action in Eq.~(\ref{MLag}) is written in Euclidean space; however the analytical continuation to Minkowski space is straightforward and can be achieved by the trivial substitutions stated in Eq.~(\ref{acont}). Therefore in Eq. (\ref{AfterGrasm}) one should assume such an analytic continuation for the light-cone variable $x^-$.}
\begin{eqnarray}
&&\Gamma^{\mu\nu}[k_1, k_3] = \frac{e^2 e_f^2 }{2}\int^\infty_0\frac{dT}{T} e^{-m^2T}~tr_c \int \mathcal{D}x\int \mathcal{D}\psi \prod^4_{i=1} \int^T_0 d\tau_i ~ (\dot{x}^\mu_1 + 2i\psi^\mu_1 \psi^\rho_1 k_{1\rho}) e^{ik_1 x_1}~(\dot{x}^\nu_3 + 2i\psi^\nu_3 \psi^\sigma_3 k_{3\sigma}) e^{ik_3 x_3}
\nonumber\\
&&\times ~ \Big(\dot{x}^-_2\,{\rm sign}(\dot{x}^-_2)\, \delta(x^-_2) + i g \psi^\xi_2 \psi^\lambda_2 F_{\xi\lambda}(x_2) \Big) \Big(\dot{x}^-_4 \, {\rm sign}(\dot{x}^-_4)\, \delta(x^-_4) + i g  \psi^\eta_4 \psi^\kappa_4 F_{\eta\kappa}(x_4)\Big)~
\nonumber\\
&&\times~ \exp\Big\{-\int^T_0 d\tau \Big(\frac{1}{4} \dot{x}^2 + \frac{1}{2}\psi_\mu\dot{\psi}^\mu + ig\dot{x}^\mu A_\mu \Big)\Big\}.
\label{AfterGrasm}
\end{eqnarray}

From this expression, we see that the interaction of the CGC background with the world-line trajectory can be represented by the current
\begin{eqnarray}
j(x_i) = \dot{x}^-_i \,{\rm sign}(\dot{x}^-_i)\, \delta(x^-_i) + i g \psi^\xi_i \psi^\lambda_i F_{\xi\lambda}(x_i)\,,
\label{efcur}
\end{eqnarray}
which resembles the structure of the interaction of the world-line  current with the incoming photon in Eq. (\ref{current}). 
This form of the current is further modified once we take into account the boson $\dot{x}^\mu A_\mu$ phase factor in Eq. (\ref{AfterGrasm}). 
This phase depends explicitly on the path taken by the world-line. However due to the fact that interaction of the world-line with the CGC background shrinks to a point, the world-line path does not propagate in the transverse direction and can be approximated as a straight line along the light-cone. We can therefore approximate this phase factor as 
\begin{eqnarray}
W = \exp\Big\{-ig \int^T_0 d\tau~\dot{x}^- A^+\big(x^-(\tau), x_\perp(\tau)\big)\Big\}.
\label{Wlinegen}
\end{eqnarray}
This expression is immediately recognizable as a Wilson loop which depends on the transverse position of the world-line at the point of interaction with the CGC background field. Since the background field shrinks to a single point, the dependence of this phase factor on the world-line trajectory $x(\tau)$ is trivial. 

The phase factor in Eq.~(\ref{Wlinegen}) can be represented as a modification to the effective current in Eq.~(\ref{efcur}). If the world-line trajectory goes to plus infinity, i.e. $\dot{x}_i >0$, the phase factor modifies the current to read as
\begin{eqnarray}
j_W(x_i) = \dot{x}^-_i \delta(x^-_i)U(x_\perp) + i g \psi^\xi_i \psi^\lambda_i U_{[\infty, x]}(x_\perp)F_{\xi\lambda}(x_i)U_{[x, -\infty]}\,,
\label{efcurwithph}
\end{eqnarray}
where we introduced the notations,
\begin{eqnarray}
U_{[x,y]}(x_\perp) = \exp\Big\{-ig \int^x_y dx^- A^+(x^-, x_\perp)\Big\}\,,
\label{phasefactors}
\end{eqnarray}
and $U(x_\perp)\equiv U_{[\infty, -\infty]}(x_\perp)$.\footnote{Note that phase factors $U$ have color indices; for brevity, we do not write these out explicitly.}  Eq.~(\ref{Wlinegen}) manifests itself in the effective current as both infinite and semi-infinite Wilson lines; the two semi-infinite Wilson lines in the second term of the current correspond to the two segments of the world-line before and after interaction with the shockwave.
Likewise, if the world-line at the point of interaction $\tau_i$ goes to minus infinity, or $\dot{x}_i < 0$, the phase in Eq.~(\ref{Wlinegen}) modifies the current to read instead as 
\begin{eqnarray}
j_W(x_i) = \dot{x}^-_i \delta(x^-_i)U^\dag(x_\perp) + i g \psi^\xi_i \psi^\lambda_i U_{[-\infty, x]}(x_\perp)F_{\xi\lambda}(x_i)U_{[x, \infty]}\,.
\label{efcurwithph2}
\end{eqnarray}
While Eqs.~(\ref{efcurwithph}) and (\ref{efcurwithph2})  provide different orderings in the color trace of Eq. (\ref{AfterGrasm}), their contribution to the functional integrals over $x(\tau)$ and $\psi(\tau)$ are the same.

In the derivation of the world-line currents Eqs.~(\ref{efcurwithph}) and (\ref{efcurwithph2}), we used only the shock-wave property of the CGC background. It is important that these expressions include Wilson line factors which are defined by an infinite number of interactions with the background field. This is crucial since it represents all-twist contributions that are equally important at small $x$.  Moreover the presence of these Wilson line factors in the currents makes our final expression for the polarization tensor in Eq. (\ref{AfterGrasm2}) manifestly gauge invariant.


An important property of our result for the currents is that it includes in principle the contribution of the transverse component of the background field $A_\perp$. The inclusion of this component of the gauge field is a step beyond the eikonal approximation in the CGC and is necessary to describe spin effects at small $x$. The transition of polarization from gauge fields to the world-line fermions is given by the Grassmann term in the currents (Eqs.~(\ref{efcurwithph}) and (\ref{efcurwithph2})) with the operator structure $U_{[-\infty, x]}(x_\perp)F_{mn}(x_i)U_{[x, \infty]}$. This result is in agreement with recent studies of spin effects at small $x$ \cite{Kovchegov:2015pbl,Kovchegov:2016weo,Kovchegov:2018znm}, where the $F_{mn}$ strength tensor was introduced at  leading order in perturbation theory to ensure gauge covariance of the current. However the world-line approach allows us to directly identify the full structure of the operator which includes the contribution of all orders of expansion in the coupling constant essential at small $x$. As we showed, this structure is the consequence of the shockwave approximation of the background field on the world-line. In paper II~\cite{inprep:2019}, we will study spin effects at small $x$ using the full form of the currents in Eqs.~(\ref{efcurwithph}) and (\ref{efcurwithph2}). We will restrict ourselves here to the case of unpolarized scattering and the computation of the structure function $F_2$, for which case the form of the currents is simpler.

Indeed, the structure of Eqs.~(\ref{efcurwithph}) and (\ref{efcurwithph2}) becomes especially simple when we take into account the explicit form of the CGC classical field in Eq.~(\ref{BF}) which has $A_\perp = 0$, leaving only one nonzero component of the gauge field. In this case, one can express 
\begin{eqnarray}
&&\partial^m U(x_\perp) = -ig \int d x^- ~ U_{[\infty, x]}F^{m+}(x^-, x_\perp) U_{[x, -\infty]}\,,
\label{relation}
\end{eqnarray}
such that the the currents Eqs.~(\ref{efcurwithph}) and (\ref{efcurwithph2}) can be reexpressed as\footnote{The required integration over $x^-$ is given by the functional integral over $x(\tau)$ in Eq. (\ref{AfterGrasm}) at the point of interaction $\tau_i$.}
\begin{eqnarray}
j_W(x_i) = \big(\dot{x}^-_i + 2  \psi^-_i \psi^m_i  \partial_m \big) \delta(x^-_i) U(x_\perp)\,,
\label{efcurwithphCGC}
\end{eqnarray}
and a similar expression for the reverse time ordering of the trajectory is obtained by substituting $U\to U^\dag$. Taking into account the fact that different orderings in the color trace provides equal contributions to the functional integrals over $x(\tau)$ and $\psi(\tau)$, 
\begin{eqnarray}
&&{\rm Tr_c}~ U(x_\perp) U^\dag (y_\perp) = {\rm Tr_c}~ U^\dag(x_\perp) U(y_\perp)\,,
\end{eqnarray}
we can finally rewrite Eq. (\ref{AfterGrasm}) as
\begin{eqnarray}
&&\Gamma^{\mu\nu}[k_1, k_3] = - \frac{e^2 e_f^2 }{2}\int^\infty_0\frac{dT}{T} e^{-m^2T}~{\rm Tr_c} \int \mathcal{D}x\int \mathcal{D}\psi \prod^4_{i=1} \int^T_0 d\tau_i ~ (\dot{x}^\mu_1 + 2i\psi^\mu_1 \psi^\rho_1 k_{1\rho}) e^{ik_1 x_1}~(\dot{x}^\nu_3 + 2i\psi^\nu_3 \psi^\sigma_3 k_{3\sigma}) e^{ik_3 x_3}
\nonumber\\
&&\times ~ \big(\dot{x}^-_2 + 2  \psi^-_2 \psi^m_2  \partial_{2m} \big) \delta(x^-_2) U(x_{2\perp})~ \big(\dot{x}^-_4 + 2 \psi^-_4 \psi^n_4 \partial_{4n} \big) \delta(x^-_4) U^\dag (x_{4\perp})~\exp\Big\{-\int^T_0 d\tau \Big(\frac{1}{4} \dot{x}^2 + \frac{1}{2}\psi_\mu\dot{\psi}^\mu \Big)\Big\}.
\label{AfterGrasm2}
\end{eqnarray}

Our approximation for the scalar phase factor is only to leading order in the eikonal expansion. One can try to calculate the subleading corrections $ \int dx^i A_i$; these will however 
only matter in situations where the finite size of the background field is relevant, as is the case at large $x$. In this study, we will restrict ourselves to the high energy small $x$ limit where 
such corrections are power suppressed in the energy. With this caveat, we have everything we need to compute the hadron tensor given in Eq. (\ref{TprodEffex}).

However before we do so, we must address the violation of translational invariance introduced by the shockwave approximation. Indeed, in the derivation of Eq.~(\ref{AfterGrasm2}), we assumed that the background field is localized at the longitudinal coordinate $x^- = 0$, explicitly breaking translational invariance in the longitudinal $x^-$ direction. As proposed in \cite{McLerran:1998nk}, this can be restored by an additional integral over $X^-$, giving the formal expression\footnote{In writing this expression, we have made use of the optical theorem relating the imaginary part of the time ordered product of currents (proportional to the r.h.s) to the corresponding Wightman function on the l.h.s, which is the hadron tensor. We also made use of the fact that the normalization 
$\langle P|P\rangle$ introduces a volume factor, which in addition to $X^-$ also introduces the transverse radius $\sigma$.}
\begin{eqnarray}
W^{\mu\nu}(q, P, S) = \frac{\sigma P^+}{\pi e^2}\,{\rm Im}\  \int dX^- \int d^4x ~e^{iqx}\langle P,S| \frac{\delta^2 \Gamma[A]}{\delta A_\mu(\frac{x}{2} + X^-)\delta A_\nu(-\frac{x}{2} + X^-)} |P,S\rangle\,,
\label{TprodEffexUp}
\end{eqnarray}
where $P^+\to\infty$ is the light-cone momentum of the target and $\sigma$ is the transverse radius of the target. This can be equivalently expressed in Euclidean metric as 
\begin{eqnarray}
&&W^{\mu\nu}(q, P, S) 
\nonumber\\
&&= \frac{ \sigma P^+}{\pi e^2}\,{\rm Im}\ \int dX^- \int d^4x \,e^{-iqx} 
 \int \frac{d^4k_1}{(2\pi)^4} \int \frac{d^4k_3}{(2\pi)^4} e^{-ik_1 (\frac{x}{2} + X^- )} e^{-ik_3( - \frac{x}{2} + X^-) }\langle P,S| \Gamma^{\mu\nu}[k_1, k_3] |P,S\rangle.
\label{WModFu}
\end{eqnarray}
where $\Gamma^{\mu\nu}[k_1, k_3]$ is defined by Eq. (\ref{AfterGrasm2}). In performing this substitution, we observe that the resulting expression contains matrix elements of the product of  $U(x_{2\perp})$ and $U^\dag(x_{4\perp})$. If the matrix element is translationally invariant in the transverse direction (as for a very large nucleus), the result will depend only on the difference of the transverse coordinates $x_{2\perp}$ and $x_{4\perp}$. It is therefore convenient to introduce the function $\gamma(k_\perp)$ which satisfies 
\begin{eqnarray}
\frac{1}{N_c}{\rm Tr}_c\langle P,S| U(x_{2\perp}) U^\dag(x_{4\perp}) |P,S\rangle =  \int \frac{ d^2 k_\perp}{(2\pi)^2} e^{ ik_\perp(x_{2\perp} - x_{4\perp})} \,\gamma(k_\perp)\,.
\label{F1}
\end{eqnarray}

With this substitution, we obtain the world-line representation of the hadron tensor at small $x$ to be 
\begin{eqnarray}
&&W^{\mu\nu}(q, P, S) 
\nonumber\\
&&= - \frac{ e_f^2 \sigma P^+ N_c}{8 \pi}\,{\rm Im}\ \int \frac{ dk^+ d^2 k_\perp}{(2\pi)^3}~ \gamma(k_\perp) \int^\infty_0\frac{dT}{T} e^{-m^2T}~\int \mathcal{D}y\int \mathcal{D}\psi \prod^4_{i=1} \int^T_0 d\tau_i~\exp\Big\{-\int^T_0 d\tau \Big(\frac{1}{4} \dot{y}^2 + \frac{1}{2}\psi_\mu\dot{\psi}^\mu \Big)\Big\}
 \nonumber\\
 && ~ \Big\{(\dot{y}^\mu_1 - 2i\psi^\mu_1 \psi^\rho_1 q_\rho) e^{-iq y_1}~\big(\dot{y}^-_2 + 2 i \psi^-_2 \psi^m_2  k_{m} \big) e^{ik y_2}~(\dot{y}^\nu_3 + 2i \psi^\nu_3 \psi^\sigma_3   q_\sigma ) e^{iq y_3} ~ \big(\dot{y}^-_4 - 2 i \psi^-_4 \psi^n_4  k_{n} \big) e^{-ik y_4} \Big\}~  \Big|_{k^- = 0}\,.
 \label{WModFu2}
\end{eqnarray}
In writing this expression\footnote{We promoted the 2-vector $k_\perp$ to a four vector in the currents by replacing the delta function in $x^-$ by its equivalent integral representation, and by 
restricting $k^-=0$.}, we integrated over the zero mode in the functional integral and evaluated several coordinate and momentum integrals. In particular, we identified $k$ as the momentum transferred from the CGC background to the world-line.  Note that translational invariance is also fully restored. Further, as the factor $e_f^2$ indicates, this is the expression for a single flavor of 
quark, whose mass we have for convenience denoted as $m$ here. The final expression should be summed over all quark flavors. 
Note further that the current representing the interaction of the world-line with the shockwave background seen here to be 
\begin{eqnarray}
j(y_i) = \dot{y}^-_i + 2 i \psi^-_i \psi^m_i k_m \,,
\label{CGCcur}
\end{eqnarray}
is similar to the structure of the current in Eq.~(\ref{current}) describing the coupling of the external photon to the world-line. 

In Appendix \ref{app:5}, we will show that the world-line current $j^\mu$ corresponds to a $\gamma^\mu$ insertion in the standard Feynman diagram approach. From the form of the world-line current here, one can infer that the effective vertex of this interaction in the standard technique should be defined by an insertion of $\gamma^-$. This coincides with the conclusion of  \cite{McLerran:1998nk}, where the structure of the quark propagator in the small $x$ shockwave background was analyzed.
%

\section{World-line derivation of the dipole model at small $x$}
We have now developed all the necessary tools in the world-line formalism to compute the hadron tensor $W^{\mu\nu}$ in the small $x$ shockwave approximation. We 
will here apply these techniques to compute the  unpolarized structure functions $F_1$ and $F_2$ using the world-line representation of $W^{\mu\nu}$ in Eq.~(\ref{WModFu2}) and the projectors defined in Eq. (\ref{Proj}). 
In analogy to Eq. (\ref{PolarTerms}), we can rewrite Eq. (\ref{WModFu2}) as the sum over terms with different structures of the world-line integrals:
\begin{eqnarray}
&&W^{\mu\nu}(q, P, S) = - \frac{ e_f^2 \sigma P^+ N_c}{8\pi }\,{\rm Im}\  \int \frac{ dk^+ d^2 k_\perp}{(2\pi)^3}\,\gamma(k_\perp)\,\int^\infty_0\frac{dT}{T} e^{-m^2T}~ \prod^4_{i=1} \int^T_0 d\tau_i
\nonumber\\
&&\times \sum^{12}_{j=1} ~\int \mathcal{D}y \int \mathcal{D}\psi ~ \Delta^{\mu\nu}_j~ e^{-iq y_1}e^{ i k y_2}e^{ iq y_3}e^{ - i k y_4} ~ \exp\Big\{-\int^T_0 d\tau \Big(\frac{1}{4} \dot{y}^2 + \frac{1}{2}\psi \dot{\psi} \Big)\Big\}  \Big|_{k^- = 0}\,,
\label{WModFu3}
\end{eqnarray}
where the structure of $\Delta^{\mu\nu}_j$ is similar to Eq. (\ref{Delt1}) and can be easily reconstructed from Eq. (\ref{WModFu2}), namely, 
\begin{eqnarray}
\Delta^{\mu\nu}_1 = \dot{y}^\mu_1~ \dot{y}^-_2~ \dot{y}^\nu_3~ \dot{y}^-_4;\ \ \ \Delta^{\mu\nu}_2 = \dot{y}^\mu_1~ (2 i  \psi^-_2 \psi^m_2 k_m)~\dot{y}^\nu_3~(- 2 i \psi^-_4 \psi^n_4 k_n)\,, \ \ {\rm etc.}
\end{eqnarray}
The identity 
\begin{eqnarray}
&&\int \mathcal{D}\psi~ \psi^\mu_i\psi^\nu_i~ \exp\Big\{-\int^T_0 d\tau  \frac{1}{2}\psi \dot{\psi} \Big\} = 0\,,
\end{eqnarray}
shows that four terms in the product of brackets in Eq. (\ref{WModFu2}), such as for instance $\dot{y}^\mu_1 ~ \dot{y}^-_2~ \dot{y}^\nu_3~ (- 2 i \psi^-_4 \psi^n_4 k_n) $, do not survive leaving us
with only twelve terms in Eq. (\ref{WModFu3}).

The computation of the remaining functional integrals in Eq. (\ref{WModFu3}) can be performed using the world-line Green's function techniques presented in Appendices \ref{app:2} and \ref{app:3}. 
We will furthermore employ a momentum space representation of world-line functional integrals that is outlined in Appendices \ref{app:4} and \ref{app:5}.
 We observe that there are six different orderings of the proper time variables $\tau_i$, which can be interpreted as the independent flow of momenta in the usual language of Feynman diagrams. As an example, the structure of denominator in Eq. (\ref{fourc}) of Appendix \ref{app:5} corresponds to the functional integral over the trajectory $y(\tau)$ with the time ordering $\tau_1<\tau_2<\tau_3<\tau_4$.

Despite their apparent complexity, the six contributions with different orderings of the proper time variables $\tau_i$ in Eq. (\ref{WModFu3}) can be summarized by the two diagrams in Fig. \ref{fig:3}. All other diagrams are obtained by a simple change of variables in the corresponding momentum representation of the functional integral.
\begin{figure}[htb]
 \begin{center}
 \includegraphics[width=120mm]{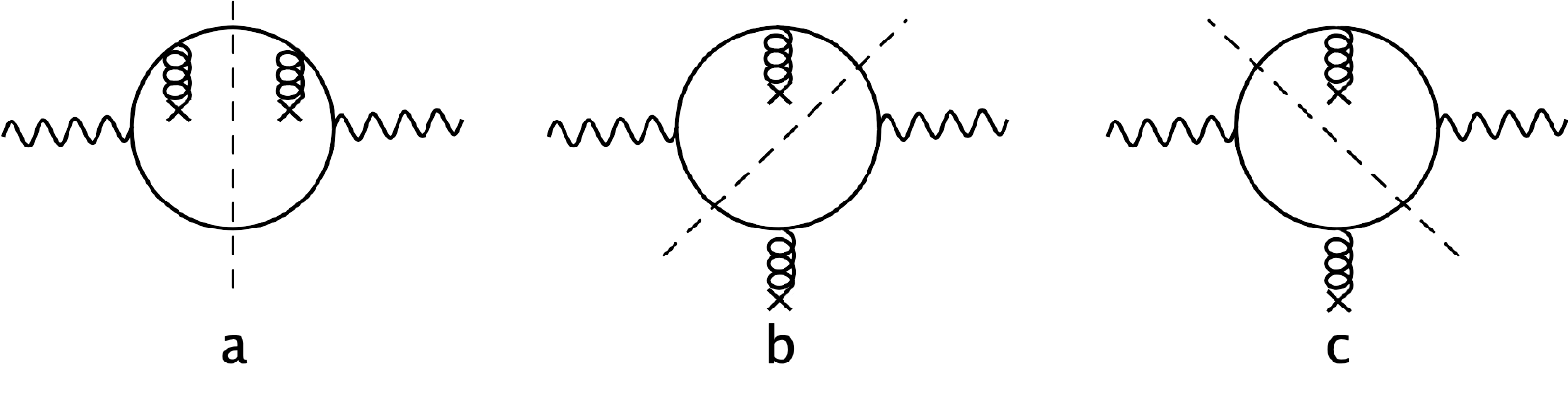}
 \end{center}
 \caption{\label{fig:4}Functional integrals with different cuts}
 \end{figure}
 In particular, using Eq.~(\ref{fourc}), the ``scalar" contribution $\Delta^{\mu\nu}_1$ in the integrand of Eq.~(\ref{WModFu3}) can be expressed as (after continuation to the Minkovski space with metric $g = (1, -1, -1, -1)$),
\begin{eqnarray}
&&W^{\mu\nu}(q, P, S)\Big|_{\Delta^{\mu\nu}_1;~\tau_1<\tau_2<\tau_3<\tau_4} =  \frac{ e_f^2 \sigma P^+ N_c}{2 \pi }\,{\rm Im}\ i \int \frac{ dk^+ d^2 k_\perp}{(2\pi)^3}\,\gamma(k_\perp)\,
\nonumber\\
&&\times ~ \int \frac{d^4 p}{(2\pi)^4} \frac{ \mathcal{N}^{\mu\nu}(p,q) }{(p^2 - m^2 + i\epsilon)((p-q)^2 - m^2 + i\epsilon)((p-q - k)^2 - m^2 + i\epsilon)((p-k)^2 - m^2 + i\epsilon)}  \Big|_{k^- = 0},
\label{WModFuDelta1}
\end{eqnarray}
where
\begin{eqnarray}
&&\mathcal{N}^{\mu\nu}(p,q) = \big(2p - 2k -q\big)^\mu\big(2p-2q - k\big)^- \big(2p - q \big)^\nu\big(2p-k\big)^- 
\nonumber\\
&&- g^{\mu-}\big((p-k-q)^2 - m^2\big)\big(2p - q \big)^\nu\big(2p-k\big)^- - g^{-\nu}\big((p-q )^2 - m^2\big)\big(2p - 2k - q\big)^\mu\big(2p-k\big)^-
\nonumber\\
&&- g^{-\nu}\big(p^2 - m^2\big)\big(2p-2k-q\big)^\mu\big(2p-2q - k\big)^- - g^{\mu-}\big((p-k)^2 - m^2\big)\big(2p-2q - k\big)^-\big(2p - q \big)^\nu
\nonumber\\
&&+ g^{\mu-}g^{-\nu}\big((p-k-q)^2 - m^2\big)\big(p^2 - m^2\big) + g^{\mu-}g^{-\nu}\big((p-k)^2 - m^2\big)\big((p-q)^2 - m^2\big)\,.
\label{fourcnumDelta_1}
\end{eqnarray}

From the structure of Eq.~(\ref{WModFuDelta1}), the calculation of the imaginary part of world-line functional integrals is straightforward, and can be done by the application of Cutkosky rules--which correspond to cuts of the world-line diagrams in Fig. \ref{fig:3}. Since for DIS we have $q^2<0$, the diagram in Fig. \ref{fig:3}a has only one cut. In contrast, the diagram in Fig. \ref{fig:3}b has two cuts, which are shown in Figs. \ref{fig:4}b and \ref{fig:4}c. In total, the six world-line diagrams with different time orderings generate eight diagrams with cuts. However only two of them have unique topologies--these correspond to the contributions of the diagrams in Fig. \ref{fig:4}a and \ref{fig:4}b. 

In particular, the diagram in Fig. \ref{fig:4}b yields,
\begin{eqnarray}
&&W^{\mu\nu}(q, P, S)\Big|_{\Delta^{\mu\nu}_1; {\rm Fig.~\ref{fig:4}b}} = -\frac{ e_f^2 \sigma P^+ N_c}{ 4 \pi } \int \frac{ dk^+ d^2 k_\perp}{(2\pi)^3}\, \gamma(k_\perp)
\nonumber\\
&&\times ~ \int \frac{d^4 p}{(2\pi)^4} \frac{ \mathcal{N}^{\mu\nu}(p,q) }{(p^2 - m^2 )((p-q - k)^2 - m^2 )} (2\pi )^2\delta((p-q)^2 - m^2)\delta((p-k)^2 - m^2)\theta(p^-)\theta(q^- - p^-)  \Big|_{k^- = 0}\,.
\label{WModFuDelta1}
\end{eqnarray}
To calculate the corresponding contribution to the structure functions, one should multiply Eq. (\ref{WModFuDelta1}) by the projectors given in Eq.~(\ref{Proj}).

A similar calculation can be done for the other terms $\Delta_j$ in Eq. (\ref{WModFu3}). If we sum all of them, all terms contributing to the topology in Fig.~\ref{fig:4}b give 
\begin{eqnarray}
&&F_2\Big|_{{\rm Fig.~\ref{fig:4}b}}
= - \frac{ e_f^2 \sigma N_c Q^2 q^-}{8 \pi } \int \frac{ dk^+ d^2 k_\perp}{2\pi}\,\gamma(k_\perp)\,\int \frac{d^4 p}{(2\pi)^4}  \theta(p^-)\theta(q^- - p^-)
\label{WModFuAllDelta_g}\\
&&\times ~  \frac{24\,Q^2  z^2 (1 - z)^2 - 2 p^2_\perp + 2 p_i k_i - 2 ( 2z - 1 )^2 m^2   - ( 4z^2 - 4z + 3 ) k^2_\perp - 6 z (1 - z ) Q^2 }{(p^2 - m^2 )((p-q - k)^2 - m^2 )} \delta((p-q)^2 - m^2)\delta((p-k)^2 - m^2)  \Big|_{k^- = 0}
\nonumber
\end{eqnarray}
where we introduced the variable $z = p^- / q^-$.

Integrating over the variables $k^+$ and $p^+$, we finally get the following contribution of Fig.~\ref{fig:4}b to the structure function $F_2$:
\begin{eqnarray}
&&F_2\Big|_{{\rm Fig.~\ref{fig:4}b}} = - \frac{e_f^2 \sigma Q^2 N_c}{64\pi^2} \int^1_0 dz \int\frac{d^2p_\perp}{(2\pi)^2}\frac{d^2 k_\perp}{(2\pi)^2} \gamma(k_\perp)
 \frac{1}{p^2_\perp + m^2 + z(1 - z)Q^2}\frac{1}{(p-k)^2_\perp + m^2 + z (1-z) Q^2} 
\nonumber\\
&&\times \Big[24\,Q^2  z^2 (1 - z)^2 - 2 p^2_\perp + 2 p_i k_i - 2 ( 2z - 1 )^2 m^2   - ( 4z^2 - 4z + 3 ) k^2_\perp - 6 z ( 1 - z ) Q^2\Big]
\label{semifinal}
\end{eqnarray}
where, recall that  
\begin{eqnarray}
\gamma(k_\perp) = \frac{1}{N_c}\int d^2x_\perp e^{-ik_\perp x_\perp} {\rm Tr}_c \langle P,S| U_{x} U^\dag_{0} |P,S\rangle\, .
\label{defg2}
\end{eqnarray}
The contribution of the second nontrivial contribution represented by the  cut in Fig.~\ref{fig:4}a coincides with Eq. (\ref{semifinal}), except with a different common sign and setting $k_\perp = 0$ in all the terms in the integrand multiplying the function $\gamma(k_\perp)$.

Taking sum of contributions in Figs.~\ref{fig:4}a and \ref{fig:4}b we find
\begin{eqnarray}
&&F_2\Big|_{{\rm Fig.~\ref{fig:4}a~+~Fig.~\ref{fig:4}b}} = - \frac{e_f^2 \sigma Q^2 N_c}{16\pi^2}\int^1_0 dz \int\frac{d^2p_\perp}{(2\pi)^2}\frac{d^2 k_\perp}{(2\pi)^2} \tilde{\gamma}(k_\perp)
 \frac{1}{p^2_\perp + m^2 + z(1 - z)Q^2}\frac{1}{(p-k)^2_\perp + m^2 + z (1-z) Q^2} 
\nonumber\\
&&\times \Big[4 z^2(1 -  z )^2 Q^2 + m^2 + \big( z^2 + (1-z)^2 \big) p_\perp\cdot (p-k )_\perp\Big]\,.
\label{final1}
\end{eqnarray}
Here we defined 
\begin{eqnarray}
\tilde{\gamma}(k_\perp) = \frac{1}{ N_c}\int d^2x_\perp e^{-ik_\perp x_\perp} {\rm Tr}_c\langle P,S| U_{x} U^\dag_{0} - 1 |P,S\rangle\,.
\end{eqnarray}

We can perform the integration over the transverse momenta $p_\perp$ in all the terms using the identities,
\begin{eqnarray}
\int d^2p_\perp \frac{1}{(p^2_\perp + \epsilon^2)((p-k)^2_\perp + \epsilon^2)} = \int d^2r_\perp e^{ik_\perp r_\perp}K^2_0(\epsilon r_\perp)\,,
\end{eqnarray}
and
\begin{eqnarray}
&&\int d^2p_\perp  \frac{p_\perp\cdot (p-k)_\perp}{(p^2_\perp + \epsilon^2)((p-k)^2_\perp + \epsilon^2)} = \epsilon^2\int d^2r_\perp  e^{ik_\perp r_\perp} K^2_1(\epsilon r_\perp)\,,
\end{eqnarray}
where $\epsilon^2 = m^2 + Q^2z(1 - z)$. Then taking the sum over all terms, we obtain the final result,
\begin{eqnarray}
&&F_2 = \frac{\sigma Q^2 N_c}{2\pi^3} \sum_f e_f^2 \int^1_0 dz \int dr_\perp~r_\perp ~\big( 1 -  \gamma(r_\perp) \big) 
\nonumber\\
&&\times~\Big[ \big(4Q^2z^2(1 - z)^2 \big)  K^2_0(\epsilon_f r_\perp) + \left[\big(z^2 + (1-z)^2\big) \epsilon_f^2 K^2_1(\epsilon_f r_\perp)+m_f^2  K^2_0(\epsilon_f r_\perp)\right]\Big]\,,
\label{final}
\end{eqnarray}
where the survival probability $\gamma(r_\perp)$ is the Fourier transformation of Eq.~(\ref{defg2}). We have also summed over all quark flavors, replacing in the process $m\rightarrow m_f$ and $\epsilon\rightarrow \epsilon_f$.  

The first and second terms in the brackets are proportional to the probabilities respectively for a longitudinal and transversely polarized photon to split into a $q\bar{q}$ pair given by the splitting functions~\cite{Bjorken:1970ah}
\begin{eqnarray}
&&|\Psi_L^f(z, r)|^2=\frac{\alpha_{\rm em} N_c e_f^2 }{2\pi^2}~ 4Q^2z^2(1-z)^2K^2_0(\epsilon_f r)\,,
\nonumber\\
&&|\Psi_T^f(z, r)|^2=\frac{\alpha_{\rm em} N_c e_f^2}{2\pi^2}~\left[\big(z^2 + (1-z)^2\big) \epsilon_f^2 K^2_1(\epsilon_f r)+m_f^2 K^2_0(\epsilon_f r)\right]\,,
\end{eqnarray}
where $\alpha_{\rm em}$ is the QED fine structure constant. Thus $F_2\propto (|\Psi_L(z, r)|^2 + |\Psi_T(z, r)|^2)$, the sum of the probabilities of transversally and longitudinally polarized photons to split into a quark-anti-quark pair. In Eq. (\ref{final}), one can therefore recognize the standard dipole model form\footnote{Note that $|\Psi_{L,T}^f(z, r)|^2$ is defined slightly different in Ref. \cite{Kovchegov:2012mbw} with the difference absorbed in a redefinition of the phase space measure.} of the structure function at small $x$~\cite{Nikolaev:1990ja,Mueller:1989st}. We thus see that the world-line representation of the hadron tensor properly captures the small $x$ dynamics of a $q\bar{q}$ interacting with the shockwave background. 

The longitudinally polarized structure function $F_L$, which can be independently extracted in DIS by varying the electron energy for a given $Q^2$ and Bjorken $x$, is obtained by simply replacing the sum of the two splitting probabilities in Eq.~(\ref{final}) with the probability of a longitudinally polarized photon to split into a quark-antiquark pair -- the term proportional to $K_0^2$. 
Using the definition $2x\,F_1 = (F_2 -F_L)$, we can obtain equivalently the structure function $F_1$.

As a final comment, recall that the world-line current has the form
\begin{eqnarray}
j(x_i) = \dot{x}^-_i \delta(x^-_i) + i g \psi^\xi_i \psi^\lambda_i F_{\xi\lambda}(x_i)\,.
\label{fcurrent}
\end{eqnarray}
This current transparently indicates how the spin of the world-line couples to the background field from the target. As we will show in our follow-up paper, this form of the current can be employed to compute the spin-dependent structure function $g_1$.

\section{Conclusions}
In this paper, we developed a world-line formalism to compute structure functions in DIS at small $x$. Starting from the expression relating the hadron tensor in DIS to the time ordered product of currents, we rewrote the latter in terms of world-line path integrals. We discussed the simpler examples of scalar and spinor QED, as well as the vacuum polarization tensor, before discussing the QCD case. For the latter, we computed the polarization tensor in the gluon shockwave background, which provides the leading contribution in the CGC EFT at small $x$. We showed how one extracts from this expression the well-known dipole model expression for the unpolarized structure function $F_2$. In doing so, we established a dictionary between computations in the 
world-line framework to that of Feynman diagrams. 

The techniques developed here will can be extended to the case of polarized structure functions at small $x$~\cite{inprep:2019} and to explore the role of the chiral anomaly in such experiments. Further, since the world-line formalism provides a natural framework to describe phase-space Wigner distributions~\cite{Mueller:2019gjj}, it also provides an ab initio framework to compute both one-dimensional helicity distributions, as well as more differential questions regarding the distributions of partons in both momentum and impact parameter space~\cite{Burkardt:2002hr,Belitsky:2003nz,Lorce:2011ni}. These will be particularly interesting in light of forthcoming experiments at a polarized Electron-Ion collider~\cite{Accardi:2012qut,Aschenauer:2017jsk}.  More speculatively, an interesting possibility is that of formulating worldline computations of structure functions and Wigner distributions in the Regge limit as a hybrid quantum computational problem. Work in this direction will be reported separately~\cite{inprepQC:2019}. 

\section*{Acknowledgements}
We gratefully acknowledge useful and inspiring conversations with Jochen Bartels, Yuri Kovchegov, Niklas Mueller, Barbara Pasquini, Daniel Pitonyak, Matt Sievert and Feng Yuan. A.~T. and R.~V.'s work is supported by the U.S. Department of Energy, Office of Science, Office of Nuclear Physics, under Contracts No. DE-SC0012704 and within the framework of  the TMD Topical Theory Collaboration. Their work is also supported in part by an LDRD grant from Brookhaven Science Associates. 
\appendix

\section{World-line representation of the effective action for spinor QED 
\label{app:1}}
Let us consider the more complicated case of spinor QED  defined by the Lagrangian
\begin{eqnarray}
\mathcal{L}_{QED} = \bar{\psi}(i\slashed{\partial} - e\slashed{A})\psi - m^2 \bar{\psi} \psi\,.
\end{eqnarray}
In  full analogy with the scalar case, one obtains the following representation of the effective action of spinor QED:
\begin{eqnarray}
\Gamma_{QED}[A] = -\frac{1}{2} {\rm Tr}\ \int^\infty_0 \frac{dT}{T} e^{-Tm^2}\exp\Big\{- T \Big[ ( p_\mu + e A_\mu)^2 - \frac{e}{2} F_{\mu\nu} \sigma^{\mu\nu}\Big] \Big\}\,,
\label{QEDheatkr}
\end{eqnarray}
where we used the identity
\begin{eqnarray}
 (\slashed{p} + e\slashed{A})^2 + m^2 = ( p_\mu + e A_\mu)^2 - \frac{e}{2} F_{\mu\nu} \sigma^{\mu\nu} + m^2\,,
\end{eqnarray}
defining the antisymmetric matrix $\sigma^{\mu\nu} = \frac{i}{2}[\gamma^\mu, \gamma^\nu]$.

Comparing the effective actions in Eqs.~(\ref{EA}) and  (\ref{QEDheatkr}), one concludes that the boson and fermion sectors of spinor QED are separable. The structure of the bosonic component is identical to the case of the scalar QED, while the quark's spinor structure is described by the $F_{\mu\nu} \sigma^{\mu\nu}$ term.
We will now follow a similar procedure to the scalar QED case to construct the world-line functional integral representation of the spinor part of the QED effective action. However to do this, one has to introduce fermionic coherent states similar to the complete set of coordinate and momentum bosonic coherent states $|x\rangle$ and $|p\rangle$ introduced in the scalar QED case. 
A nice discussion of such a construction can be found in  \cite{Ohnuki:1978jv}; see also \cite{DHoker:1995uyv,DHoker:1995aat} for an alternative realization.

We first construct the fermion raising and lower operators $a^+_i$ and $a^-_i$ respectively, defined by~\cite{Schubert:2001he}
\begin{eqnarray}
a^{\pm}_i \equiv \frac{1}{2}(\gamma_i \pm i\gamma_{i+2}),\ \ \ i = 1,\ 2.
\end{eqnarray}
where $\gamma_i$ are Euclidean gamma matrices defined by the identity $\{\gamma_i, \gamma_j\} = 2g_{ij}$.
One can check that these operators satisfy the anticommutation relations
\begin{eqnarray}
\{a^+_i, a^-_j\} = \delta_{ij},\ \ \ \{a^+_i, a^+_j\} = \{a^-_i, a^-_j\} = 0
\end{eqnarray}
With these definitions, one can introduce a Hilbert space with a vacuum defined by
\begin{eqnarray}
a^-_r|0\rangle = 0;\ \ \ \langle  0| a^+_r = 0 \,.
\end{eqnarray}
The complete set of coherent states which define the basis of the Hilbert space can be obtained by action of the raising and lowering operators on the vacuum. Such states should satisfy
\begin{eqnarray}
\langle \xi | a^-_i = \langle \xi | \xi_i,\ \ \ a^-_i|\xi\rangle = \xi_i |\xi\rangle, \ \ \ \langle \bar{\xi} | a^+_i = \langle \bar{\xi} | \bar{\xi}_i,\ \ \ a^+_i |\bar{\xi}\rangle = \bar{\xi}_i |\bar{\xi}\rangle
\label{cohstdef}
\end{eqnarray}
Each fermionic coherent state $|\xi\rangle$ and $|\bar{\xi}\rangle$ is characterized by a set of Grassmann numbers $\xi_i$ and $\bar{\xi}_i$, which are defined by $\{\xi_i, \xi_j\} = 0$ and 
$\left\{\bar{\xi}_i,\bar{\xi}_j\right\}=0$. The integration over Grassmann numbers can be done with the help of the identities~\cite{Berezin:1966nc,Ohnuki:1978jv}: 
\begin{eqnarray}
&&\int d\xi_i = \int d\bar{\xi}_i = 0\,;\ \  \int \xi_i d\xi_i = \int \bar{\xi}_i d\bar{\xi}_i = i\,;\ \ \int \delta(\xi, \xi') d\xi = 1\,;\ \ \int \xi \delta(\xi, \xi') d\xi = \xi',
\end{eqnarray} 
where the delta function for Grassmann numbers is $\delta(\xi, \xi') \equiv \frac{1}{i}(\xi - \xi')$.

The fermionic coherent states which satisfy Eq.~(\ref{cohstdef}) can be explicitly constructed:
\begin{eqnarray}
\langle \xi | = -i \langle 0| \prod^2_{i=1} \delta(\xi_i, a^-_i), \ \ \ | \xi \rangle = e^{- \sum^2_{i=1}\xi_i a^+_i}|0 \rangle,\ \ \ \langle\bar{\xi}| = \langle 0| e^{- \sum^2_{i=1} a^-_i \bar{\xi}_i },\ \ \ |\bar{\xi}\rangle = -i \prod^2_{i=1}\delta(\bar{\xi}_i, a^+_i)|0\rangle 
\label{cohstateform}
\end{eqnarray}

The fermionic coherent states $|\xi\rangle$ and $|\bar{\xi}\rangle$ are analogous to the bosonic states $|x\rangle$ and $|p\rangle$ and satisfy similar completeness and orthogonality relations, see (\ref{bosonort}):
\begin{eqnarray}
i \int |\xi\rangle \langle \xi| d^2\xi = -i \int |\bar{\xi}\rangle \langle \bar{\xi}| d^2\bar{\xi} = 1\,;\ \ \ \langle\xi|\bar{\xi}\rangle = e^{\sum^2_{i=1}\xi_i\bar{\xi}_i}\,;\ \ \ \langle\bar{\xi}|\xi\rangle = e^{\sum^2_{i=1}\bar{\xi}_i \xi_i}\,,
\label{comprelferm}
\end{eqnarray}
where $d^2\xi = d\xi_2d\xi_1$, $d^2\bar{\xi} = d\bar{\xi}_1 d\bar{\xi}_2$.
 
With the complete set of states given in Eq.~(\ref{cohstateform}), the derivation of the functional integral representation of the effective action in Eq.~(\ref{QEDheatkr}) is straightforward~\cite{Schubert:2001he}. We start with the expression defining the fermionic functional trace similar to Eq. (\ref{bosontrace}):
\begin{eqnarray}
{\rm Tr}\ \mathcal{O} = i \int d^2\xi\langle -\xi|\mathcal{O}|\xi\rangle \,,
\end{eqnarray}
where
\begin{eqnarray}
\langle -\xi | = -i \int~ e^{\sum^2_{i=1}\bar{\xi}_i \xi_i}\langle \bar{\xi}| d^2\bar{\xi}\,.
\end{eqnarray}

Employing these relations brings us to the form of the effective action in Eq.~(\ref{QEDheatkr}):
\begin{eqnarray}
\Gamma_{QED}[A] = - \frac{i}{2}\ \int^\infty_0 \frac{dT}{T} e^{-Tm^2}\int d^4x \int d^2\xi\ \langle x, -\xi|\exp\Big\{- T \Big[ ( p + e A )^2 - \frac{e}{2} F_{\mu\nu} \sigma^{\mu\nu}\Big] \Big\} |x,\xi\rangle,
\end{eqnarray}
where for the bosonic functional trace we repeat steps similar to the scalar case. Splitting the integration over $T$ into $N$ segments and using completeness relations in (\ref{bosonort}) and (\ref{comprelferm}) we get
\begin{eqnarray}
\Gamma_{QED}[A] = - \frac{i^N}{2}\ \int^\infty_0 \frac{dT}{T} e^{-Tm^2} \int_{BC} \prod^N_{i=1} d^4x_i d^2\xi_i \ \langle x_{i+1}, \xi_{i+1}|\exp\Big\{- \frac{T}{N} \Big[ ( p + e A )^2 - \frac{e}{2} F_{\mu\nu} \sigma^{\mu\nu}\Big] \Big\} |x_i, \xi_i \rangle,
\label{EAintset}
\end{eqnarray}
where integration over $x_i$ and $\xi_i$ ressatisfies periodic $x_1 = x_{N+1}$ and anti-periodic $\xi_1 = -\xi_{N+1}$ boundary conditions.

Let us consider the matrix element of the evolution operator which is sandwiched between the ``coordinate" states $|x\rangle$ and $|\xi\rangle$. Let us expand the exponent and insert a complete set of "momentum" states $|p\rangle$ and $|\bar{\xi}\rangle$ in between:
\begin{eqnarray}
&&\langle x_{i+1}, \xi_{i+1}|\exp\Big\{- \frac{T}{N} \Big[ ( p + e A )^2 - \frac{e}{2} F_{\mu\nu} \sigma^{\mu\nu}\Big] \Big\} |x_i, \xi_i \rangle
\label{matrixferm}\\
&&= \int \frac{d^4p_{i+1,i} }{(2\pi)^4} e^{i p_{i+1,i} (x_{i+1} - x_i )} \Big( 1 - (\tau_{i+1} - \tau_i) \Big\{(p_{i+1,i} + e A_{i+1,i} )^2 \langle \xi_{i+1}| \xi_i \rangle - \frac{e}{2} F_{\mu\nu} \langle \xi_{i+1}| \sigma^{\mu\nu} | \xi_i \rangle \Big\} + \dots \Big)
\nonumber\\
&&= - i \int \frac{d^4p_{i+1,i} d^2 \bar{\xi}_{i+1,i} }{(2\pi)^4} e^{i p_{i+1,i} (x_{i+1} - x_i )} e^{(\xi^k_{i+1} - \xi^k_i)\bar{\xi}^k_{i+1, i}} \Big( 1 - (\tau_{i+1} - \tau_i) \Big\{(p_{i+1,i} + e A_{i+1,i} )^2 - ie F_{\mu\nu} \, \tilde{\psi}^\mu_{i+1}\psi^\nu_i \Big\} + \dots \Big)\,,
\nonumber
\end{eqnarray}
where we used the following formula for the matrix element of the product of two gamma matrixes:
\begin{eqnarray}
\langle \xi_{i+1}|\gamma^\mu \gamma^\nu | \xi_i\rangle = - 2 i \int d^2\bar{\xi}_{i+1, i} \langle \xi_{i+1}|\bar{\xi}_{i+1, i}\rangle\langle \bar{\xi}_{i+1, i}|\xi_i\rangle\, \tilde{\psi}^\mu_{i+1} \psi^\nu_i
\end{eqnarray}
introducing a shorthand notation for the coefficients
\begin{eqnarray}
&&\psi^{1,2}_i = \frac{1}{\sqrt{2}}(\xi^{1,2}_i + \bar{\xi}^{1,2}_{i+1, i}),\ \ \ \psi^{3,4}_i = \frac{i}{\sqrt{2}}(\xi^{1,2}_i - \bar{\xi}^{1,2}_{i+1, i})
\nonumber\\
&&\tilde{\psi}^{1,2}_{i+1} = \frac{1}{\sqrt{2}}(\xi^{1,2}_{i+1} + \bar{\xi}^{1,2}_{i+1, i}),\ \ \ \tilde{\psi}^{3,4}_{i+1} = \frac{i}{\sqrt{2}}(\xi^{1,2}_{i+1} - \bar{\xi}^{1,2}_{i+1, i})
\label{newvar}
\end{eqnarray}

Next, we substitute the matrix element  in Eq.~(\ref{matrixferm}) back into the effective action, symmetrize the exponential factor, and take the limit $N\to\infty$ to obtain, 
\begin{eqnarray}
&&\Gamma_{QED}[A] = - \frac{1}{2} \int^\infty_0 \frac{dT}{T} e^{-m^2T} 
\nonumber\\
&&\times\int_{PBC} \mathcal{D}x \int \mathcal{D}p \int_{APBC} \mathcal{D}\xi \mathcal{D}\bar{\xi}    \ \mathcal{P} \exp\Big\{ \int^T_0 d\tau \Big(i p \dot{x} + \frac{1}{2} \dot{\xi} \bar{\xi} - \frac{1}{2}\xi \dot{\bar{\xi}} - (p_\mu + e A_\mu )^2 + ie F_{\mu\nu} \, \psi^\mu\psi^\nu \Big) \Big\}.
\label{EAfuncQED}
\end{eqnarray}

Finally, we rewrite the fermionic term
\begin{eqnarray}
\frac{1}{2} \dot{\xi} \bar{\xi} - \frac{1}{2}\xi \dot{\bar{\xi}} = -\frac{1}{2}\psi\cdot \dot{\psi}
\end{eqnarray}
in terms of the real components of the continuum version of the variables in Eq.~(\ref{newvar}):
\begin{eqnarray}
&&\psi_{1,2} = \frac{1}{\sqrt{2}}(\xi_{1,2} + \bar{\xi}_{1,2}),\ \ \ \psi_{3,4} = \frac{i}{\sqrt{2}}(\xi_{1,2} - \bar{\xi}_{1,2})
\end{eqnarray}

As a result, one can write down the final form of the world-line representation of the effective action of the spinor QED:
\begin{eqnarray}
&&\Gamma_{QED}[A] = -\frac{1}{2} \int^T_0 \frac{dT}{T} e^{-m^2T} 
\nonumber\\
&&\times \int_{PBC} \mathcal{D}x \int_{APBC} \mathcal{D} \psi \exp\Big\{-\int^T_0 d\tau \Big(\frac{1}{4} \dot{x}^2 + \frac{1}{2}\psi_\mu\dot{\psi}^\mu + ie\dot{x}^\mu A_\mu - ie \psi^\mu \psi^\nu F_{\mu\nu}\Big)\Big\}
\end{eqnarray}

\section{Calculation of scalar functional integrals
 \label{app:2}}
In this section, we will calculate scalar functional integrals\footnote{The results of the computation of the world-line scalar integrals is well-known \cite{Strassler:1992zr}. We will present here a different approach through the notion of the semi-classical approximation around the classical trajectory; this will be promising for future computations of world-line integrals in arbitrary backgrounds.}
\begin{eqnarray}
\int \mathcal{D}x\ e^{ik_1 x_1} e^{ik_1 x_2} e^{S_B(x) };\ \ \ \int \mathcal{D}x\ \dot{x}^{\mu_1}_1 e^{ik_1 x_1} e^{ik_1 x_2} e^{S_B(x) };\ \ \ \int \mathcal{D}x\ \dot{x}^{\mu_1}_1 \dot{x}^{\mu_2}_2 e^{ik_1 x_1} e^{ik_1 x_2} e^{S_B(x) }\,,
\label{scalarint}
\end{eqnarray}
where the free bosonic world-line action on a trajectory $x(\tau)$ is
\begin{eqnarray}
S_B(x) = - \frac{1}{4} \int^T_0 d\tau~\dot{x}^2(\tau) = \frac{1}{2}\int^T_0 d\tau d\tau'~ x(\tau)G^{-1}_B(\tau, \tau') x(\tau')\,,
\label{actionsc}
\end{eqnarray}
with the bosonic world-line propagator defined in Eq. (\ref{props}).
In the more complicated case of the interaction of world-lines with the shockwave background field, we will encounter similar integrals which have more scalar current $\dot{x}^\mu$ insertions. All of these integrals can be calculated with the method discussed here. 

Before we evaluate the functional integrals in Eq. (\ref{scalarint}), let us first take into account momentum conservation which, in the language of the world-lines, corresponds to the integration over the zero-mode. We can formally write 
\begin{eqnarray}
\int \mathcal{D}x = \int d^4x_0 \int \mathcal{D}y\,,
\label{zeroMode}
\end{eqnarray}
where the zero-mode $x_0\equiv x(0) = x(T)$ is a constant shift  $x(\tau) = y(\tau) + x_0$ in the world-line trajectory and $y(\tau)$ satisfies $y(0) = y(T) = 0$. Integration over $x_0$ then leads to the momentum conservation delta function in the integrals in Eq.~(\ref{scalarint}), which takes the form
\begin{eqnarray}
\int \mathcal{D}x~ \mathcal{I}^{\mu_1\dots\mu_N}(x)~ e^{S_B(x)} = (2\pi)^2\delta^2(k_1 + k_2) \int \mathcal{D}y~ \mathcal{I}^{\mu_1\dots\mu_N}(y)~ e^{ S_B(y) }\,,
\label{firstint1}
\end{eqnarray}
where  $\mathcal{I}^{\mu_1\dots\mu_N}(x)$ represents the integrands in Eqs. (\ref{scalarint}) and $N$ is number of factors $\dot{x}^{\mu_i}_i$.

We can formally express $\mathcal{I}^{\mu_1\dots\mu_N}(x)$ with the help of auxiliary Grassmann variables $\theta$ as
\begin{eqnarray}
\mathcal{I}^{\mu_1\dots\mu_N}(y) = \prod^N_{i=1} \int d\theta_i d\bar{\theta}_i e^{- \int^T_0 d\tau (\sum^N_{i=1} J^{\mu_i\rho} + J^\rho) y_\rho}\,,
\label{GressmannExpon}
\end{eqnarray}
where the number of auxiliary integrations $i$ is equivalent to the number of factors of the scalar current $\dot{x}(\tau)$ in the integrals\footnote{The first integral therefore has no auxiliary Grassmann integrations.} of Eq. (\ref{scalarint}).

The three currents in the integrand of Eq. (\ref{scalarint}) can then straightforwardly be replaced by the currents $J(\tau)$, which for each integral are respectively, 
 \begin{eqnarray}
&&J^\rho(\tau;\tau_1,\tau_2) = - ik^\rho_1 \delta(\tau - \tau_1) - ik^\rho_2\delta(\tau - \tau_2)\,,
\label{cur1}
\end{eqnarray}
\begin{eqnarray}
&&J^{\mu_i\rho}(\tau;\tau_i) = -g^{\mu_i\rho}\bar{\theta}_i \theta_i \delta'(\tau - \tau_i)\,.
 \label{cur3}
 \end{eqnarray}
 where the current in Eq. (\ref{cur3}) correspond to the $\dot{x}$ factors in the last two integrals of Eq. (\ref{scalarint}). Note that ``prime" in the delta-function represents its derivative with respect to the proper time.
 
 As a result of the introduction of the auxiliary integrals, the integrals in Eq. (\ref{scalarint})  can be reexpressed as 
\begin{eqnarray}
\int \mathcal{D}x~ \mathcal{I}^{\mu_1\dots\mu_N}(x)~ e^{S_B(x)} = (2\pi)^2\delta^2(k_1 + k_2) \prod^N_{i=1} \int d\theta_i d\bar{\theta}_i \int \mathcal{D}y~ e^{ S^{\mu_1\dots\mu_N}(y) }\,,
\label{funcintexp2}
\end{eqnarray}
where the world-line action is now given by\footnote{For simplicity, we will suppress the indices $\mu_1\dots\mu_N$ in $S(y)$ in the following.}
\begin{eqnarray}
S(y) = S_B(y) - \int^T_0 d\tau (\sum^N_{i=1} J^{\mu_i\rho} + J^\rho) y_\rho \,.
\end{eqnarray}
This action takes into account the world-line interaction with the background field.  In particular, the current in Eq.~(\ref{cur1}) corresponds to the interaction of the world-line with two incoming scalar particles of momenta $k_1$ and $k_2$. Eq.~(\ref{funcintexp2}) has a structure very similar to the standard functional integrals of quantum field theory, except that fields are now substituted with world-lines and currents are functions of the proper time variable $\tau$.

The functional integral in Eq. (\ref{funcintexp2}) includes the contribution of all possible world-line trajectories $y(\tau)$. However in a semi-classical picture, the dominant path is  the classical path $y_{cl}(\tau)$, given by 
\begin{eqnarray}
\frac{\delta S(y)}{\delta y_\mu}\Big|_{y = y_{cl}} = 0\,.
\label{classdef}
\end{eqnarray}
In this semi-classical approximation\footnote{A shift of variables is implied; $y(\tau)$ in Eq. (\ref{semcl}) now refers to the deviation from the classical trajectory $y_{cl}$.},
\begin{eqnarray}
\int \mathcal{D}y~ e^{ S(y) } \approx e^{S(y_{cl})} \int \mathcal{D}y~ \exp \Big[\frac{1}{2}\int^T_0 d\tau d\tau'~ y_\rho(\tau)\frac{\delta^2 S}{\delta y_\rho(\tau) \delta y_\sigma(\tau')}\Big|_{y=y_{cl}}y_\sigma(\tau')\Big]\,.
\label{semcl}
\end{eqnarray}

In general, Eq.~(\ref{classdef}) does not always allow for an analytical solution. However in the case of interactions with a finite number of external particles, the action $S(y)$ is quadratic and Eq.~(\ref{classdef}) has the straightforward solution,
\begin{eqnarray}
y^\rho_{cl}(\tau) = \int^T_0 d\tau' G_B(\tau, \tau') \big(\sum^N_{i=1} J^{\mu_i\rho}(\tau') + J^\rho(\tau')\big)\,,
\end{eqnarray}
where the classical path $y^\rho_{cl}(\tau)$ is solely defined by the external currents $J(\tau)$ and the solution in Eq.~(\ref{semcl}) is the exact result.

Here we would like to mention that in the CGC background one has to take into account an infinite number of interactions with the parton constituents of the target; one therefore  should apply the semi-classical approximation in Eq.~(\ref{semcl}), with a nontrivial equation for the classical trajectory $y(\tau)$. However as we show in the main part of the paper, due to the infinitesimal structure of the small $x$ background in the longitudinal direction, the interaction can still be described by an effective current which has a form similar to the current of a pointlike particle. 
In this shockwave limit, the semi-classical approximation to the world-line functional integrals in the CGC background provides an exact solution.

Now taking into account the fact that the second derivative of the action in Eq. (\ref{semcl}) is the inverse propagator $G^{-1}_B(\tau, \tau')$,  and using Eq.~(\ref{freefunc}), we obtain 
\begin{eqnarray}
\int \mathcal{D}y~ e^{ S(y) } = (4\pi T)^{-D/2} e^{S(y_{cl})}\,.
\label{semclres}
\end{eqnarray}
Therefore the functional integrals represented in Eq. (\ref{scalarint}) have the form
\begin{eqnarray}
\int \mathcal{D}x~ \mathcal{I}^{\mu_1\dots\mu_N}(x)~ e^{S_B(x)} = (2\pi)^2\delta^2(k_1 + k_2) (4\pi T)^{-D/2} \prod^N_{i=1} \int d\theta_i d\bar{\theta}_i e^{S(y_{cl})}\,.
\label{funcintexp3}
\end{eqnarray}
Substituting the explicit form of the currents in Eqs. (\ref{cur1} - \ref{cur3}), and integrating over the auxiliary Grassmann variables one obtains, 
\begin{eqnarray}
\int \mathcal{D}x e^{ik_1 x_1} e^{ik_2x_2}e^{S_B(x) } = (2\pi)^2\delta^2(k_1 + k_2) (4\pi T)^{-D/2} e^{k_1\cdot k_2G_B(\tau_1, \tau_2)}\,,
\label{scal1} 
\end{eqnarray}
\begin{eqnarray}
\int \mathcal{D}x\ \dot{x}^{\mu_1}_1 e^{ik_1 x_1}  e^{ik_2 x_2}  e^{S_B(x) } =  (2\pi)^2\delta^2(k_1 + k_2) (4\pi T)^{-D/2} \Big[ -  ik^{\mu_1}_2 \partial_{\tau_1} G_B(\tau_1, \tau_2)  \Big] e^{ k_1\cdot k_2 G_B(\tau_1, \tau_2)}\,,
\label{scal2}
\end{eqnarray}
\begin{eqnarray}
&&\int \mathcal{D}x \dot{x}^{\mu_1}_1 \dot{x}^{\mu_2}_2 e^{ik_1 x_1} e^{ik_2x_2}  e^{S_B(x) } =  (2\pi)^2\delta^2(k_1 + k_2)(4\pi T)^{-D/2}  \Big\{- g^{\mu_1\mu_2} \partial_{\tau_1} \partial_{\tau_2} G_B(\tau_1, \tau_2)
\nonumber\\
&& + k^{\mu_1}_2 k^{\mu_2}_1 \partial_{\tau_1}G_B(\tau_1, \tau_2)
 \partial_{\tau_1}G_B(\tau_1, \tau_2)\Big\} e^{k_1\cdot k_2 G_B(\tau_1, \tau_2)} \,.
 \label{scal3}
\end{eqnarray}
The derivatives of the bosonic world-line propagator are given in Eqs. (\ref{wlpropDer}) and (\ref{wlpropDer2}).

The generalization of this computation to an integral with arbitrary number of factors $\dot{x}$ is straightforward. We will need it in our calculation of the structure functions where integrals with four scalar currents $\dot{x}$ will appear. However, a practical implementation of the scheme leads to cumbersome integrations over Grassmann variables. Moreover, a comparison of the final results for the scalar integrals Eqs.~(\ref{scal1})-(\ref{scal3}) does not provide any simple mnemonic rules that allow one to extend these results to cases where there are a  larger number of $\dot{x}$ factors--this is seen from the explicit results given in \cite{Strassler:1992zr}.

We therefore find it useful to use a different representation of the scalar functional integrals through a momentum integration which is based on the fact that the world-line action in Eq.~(\ref{actionsc}) fully represents scalar QED \cite{Strassler:1992zr, DHoker:1995uyv}. It provides a straightforward generalization to integrals with arbitrary number of incoming currents, has a direct connection to the language of standard Feynman diagrams, and is better suited for our problem of the computation of structure functions in DIS.

\section{Calculation of Grassmann functional integrals
\label{app:3}}
In this appendix, we will follow the same procedure as in Appendix \ref{app:2} to calculate the Grassmann functional integrals
\begin{eqnarray}
\int \mathcal{D}\psi~ e^{S_F(\psi)};\ \ \ \ \int \mathcal{D}\psi~ \psi^{\mu_1}_1 \psi^{\nu_1}_1~  e^{S_F(\psi)}
;\ \ \ \ \int \mathcal{D}\psi~ \psi^{\mu_1}_1 \psi^{\nu_1}_1 \psi^{\mu_2}_2 \psi^{\nu_2}_2~  e^{S_F(\psi)}\,,
\label{spinorint}
\end{eqnarray}
where the free Grassmann world-line action on a trajectory $\psi(\tau)$ is
\begin{eqnarray}
S_F(\psi) = -\int^T_0 d\tau \frac{1}{2}\psi\dot{\psi}\,.
\end{eqnarray}

This calculation is similar to calculation of bosonic functional integrals in Appendix \ref{app:2}. It is based on the notion of classical trajectories in the Grassmann sector.

Similarly to Eq. (\ref{GressmannExpon}), we will introduce the auxiliary Grassmann variables $\theta_i$ and $\bar{\theta}_i$ and exponentiate factors $\psi^{\mu_i}_i\psi^{\nu_i}_i$ in Eq. (\ref{spinorint}) as
\begin{eqnarray}
\psi^{\mu_i}_i\psi^{\nu_i}_i = \int d\theta_i d\bar{\theta}_i ~e^{-\int^T_0d\tau~\psi^\rho(\tau) \big( \delta^{\mu_i}_\rho\theta_i\delta(\tau-\tau_i)  +  \delta^{\nu_i}_\rho\bar{\theta}_i \delta(\tau - \tau_i)\big) }
\,.
\label{expongrass}
\end{eqnarray}
As a result we can rewrite functional integrals in Eq. (\ref{spinorint}) as
\begin{eqnarray}
\prod^N_{i=1} \int \mathcal{D}\psi~ \psi^{\mu_i}_i \psi^{\nu_i}_i~  e^{S_F(\psi)} = \prod^N_{i=1} \int d\theta_i d\bar{\theta}_i \int \mathcal{D}\psi~  e^{S(\psi)}\,,
\label{GrAfterExp}
\end{eqnarray}
where the Grassmann world-line action now includes terms representing the interaction with the external currents and is given by\footnote{Again, as in the previous appendix, the dependence on the indices 
$\mu_i$ and $\nu_i$ is implicit in the expressions that follow.}
\begin{eqnarray}
S(\psi) = S_F(\psi) - \int^T_0 d\tau~\psi^\rho(\tau) \sum^N_{i=1} \big( \eta^i_\rho  +  \bar{\eta}^i_\rho\big)\,,
\label{GrassmanAction}
\end{eqnarray}
and $N$ is the number of factors $\psi^{\mu_i}_i\psi^{\nu_i}_i$ in Eq.~(\ref{spinorint}). Also, we introduced here the Grassmann currents 
\begin{eqnarray}
\eta^i_\rho = \delta^{\mu_i}_\rho\theta_i\delta(\tau-\tau_i)\,;\ \ \ \bar{\eta}^i_\rho = \delta^{\nu_i}_\rho\bar{\theta}_i \delta(\tau - \tau_i)\,.
\label{GrassmCur}
\end{eqnarray}

To calculate the Grassmann functional integrals in Eq.~(\ref{GrAfterExp}), we expand the Grassmann world-line action around the classical trajectory $\psi^\mu_{cl}(\tau)$, defined by
\begin{eqnarray}
\frac{\delta S(\psi)}{\delta \psi_\mu}\Big|_{\psi = \psi_{cl}} = 0\,.
\label{classdefGr}
\end{eqnarray}
The Grassmann functional integral then takes the form
\begin{eqnarray}
\int \mathcal{D}\psi~ e^{ S(\psi) } = e^{S(\psi_{cl})} \int \mathcal{D}\psi~ \exp \Big[\frac{1}{2}\int^T_0 d\tau d\tau'~ \frac{\delta^2 S}{\delta \psi_\rho(\tau) \delta \psi_\sigma(\tau')}\Big|_{\psi=\psi_{cl}}\psi_\sigma(\tau')\psi_\rho(\tau)\Big]\,,
\label{semclGr}
\end{eqnarray}
where
\begin{eqnarray}
\frac{\delta^2 S}{\delta \psi_\rho(\tau) \delta\psi_\sigma(\tau')}\Big|_{\psi=\psi_{cl}} = -2G^{-1}_F(\tau', \tau)\,g^{\rho\sigma}\,,
\end{eqnarray}
with the fermion world-line propagator $G_F$ defined in Eq.~(\ref{wlprop}).

The world-line trajectory $\psi_{cl}(\tau)$ is the solution of Eq.~(\ref{classdefGr}) and can be expressed as 
\begin{eqnarray}
\psi^\rho_{cl}(\tau) = -\frac{1}{2}\int^T_0 d\tau' G_F(\tau, \tau')\sum^N_{i=1}\big(\eta^\rho_i(\tau') + \bar{\eta}^\rho_i(\tau')\big)\,.
\label{ClassGrassm}
\end{eqnarray}
Integrating over deviations from the classical trajectory using Eq.~(\ref{freefunc}) yields
\begin{eqnarray}
\prod^N_{i=1} \int \mathcal{D}\psi~ \psi^{\mu_i}_i \psi^{\nu_i}_i~  e^{S_F(\psi)} = 4 \prod^N_{i=1} \int d\theta_i d\bar{\theta}_i e^{S(\psi_{cl})}
\label{GrAfterExp2}
\end{eqnarray}

Substituting Eq.~ (\ref{ClassGrassm}) in Eq.~(\ref{GrassmanAction}) then gives,
\begin{eqnarray}
&&S(\psi_{cl}) = - \frac{1}{4}\int^T_0 d\tau\,d\tau'~ \sum^N_{i=1} \big( \eta^\rho_i(\tau)  +  \bar{\eta}^\rho_i(\tau)\big) G_F(\tau, \tau')\sum^N_{j=1}\big(\eta^j_\rho(\tau') + \bar{\eta}^j_\rho(\tau')\big)\,.
\label{GrassmanActionCl}
\end{eqnarray}

Further substituting the explicit form of the currents in Eqs.~(\ref{GrassmanActionCl}), and integrating over the auxiliary Grassmann variables in Eq. (\ref{GrAfterExp2}), one obtains
\begin{eqnarray}
\int \mathcal{D}\psi~  e^{S_F(\psi)} = 4\,,
\end{eqnarray}
\begin{eqnarray}
\int \mathcal{D}\psi~ \psi^{\mu_1}_1 \psi^{\nu_1}_1~  e^{S_F(\psi)} = 0\,, 
\end{eqnarray}
\begin{eqnarray}
 &&\int \mathcal{D}\psi~ \psi^{\mu_1}_1 \psi^{\nu_1}_1~ \psi^{\mu_2}_2 \psi^{\nu_2}_2 ~  e^{S_F(\psi)} = - g^{\mu_1\mu_2} g^{\nu_1\nu_2} + g^{\mu_1\nu_2} g^{\nu_1\mu_2}\,.
 \label{GrFInt3}
\end{eqnarray}
where we used the identity $G_F(\tau, \tau) = 0$.

Using the general result in Eq.~(\ref{GrAfterExp2}), one can then easily calculate functional integrals over Grassmann world-line trajectories with arbitrary number\footnote{In particular, in the main text of the paper, we use Grassmann functional integrals with $N = 3$ and $N=4$ which we calculated using Eq. (\ref{GrAfterExp2}). However since the exact formulas are rather cumbersome, we do not give them here.} $N$ of factors $\psi^{\mu_i}_i\psi^{\nu_i}_i$ in Eq. (\ref{spinorint}).

\section{World-line functional integrals in the momentum representation
\label{app:4}}
We shall introduce here a momentum representation of world-line scalar functional integrals with a finite number of interactions which in general have the form of Eq.~(\ref{semclres}). This representation is based on the simple expression for a Gaussian integral in $D$ dimensions:
\begin{eqnarray}
\int \frac{d^D p}{(2\pi)^D} e^{-T p^2} = (4\pi T)^{-D/2}\,.
\label{gauss}
\end{eqnarray}
which we shall employ to introduce momentum integrals in the semi-classical approximation for the functional integral in Eq.~(\ref{semclres}).

Before we do this, let us notice that in the computation of the polarization tensor (see for instance Eq.~(\ref{polarization})), the scalar integrals are integrated over the size of the world-line $T$ and the position of the interaction points $\tau_i$. This, in combination with Eq.~(\ref{gauss}), yields integrals with the structure
\begin{eqnarray}
\int^\infty_0 \frac{dT}{T} e^{-m^2T} \prod_i \int^T_0 d\tau_i \int \mathcal{D}y~ e^{ S(y) } = \int \frac{d^D p}{(2\pi)^D} \int^\infty_0 \frac{dT}{T} \prod_i \int^T_0 d\tau_i \,e^{-T (p^2 + m^2)}  e^{S(y_{cl})}\,,
\end{eqnarray}
where the index $i$ represents the number of interactions with external currents. Finally, one can take the explicit form of the action $S(y_{cl})$ and perform the integration over the variables $T$ and $\tau$, using the methods discussed previously.

As an example, let us consider the integral in Eq.~(\ref{scal3}). Substituting the world-line propagator and its derivatives yields,
\begin{eqnarray}
&&\int \mathcal{D}y~ \dot{y}^\mu_1 \dot{y}^\nu_2~ e^{iq y_1} e^{-iqy_2}  e^{ S_B(y) } =  \int \frac{d^D p}{(2\pi)^D} e^{-T [p^2 + q^2 u(1 - u) ]} \Big\{ \frac{2g^{\mu\nu}}{T}\delta(u) - \frac{2g^{\mu\nu}}{T} - q^\mu q^\nu (1 - 2u)^2 \Big\}
\label{mom1}
\end{eqnarray}
where $\tau_2 = uT$ and we use the rotational invariance of the world-line to fix the position of the ``first" interaction $\tau_1 = 0$. 

From Eq. (\ref{gauss}) it is easy to see that
\begin{eqnarray}
\int \frac{d^D p}{(2\pi)^D} ~\frac{g^{\mu\nu}}{T}~e^{-T p^2} = \int \frac{d^D p}{(2\pi)^D}~2p^\mu p^\nu~ e^{-T p^2}
\end{eqnarray}
Using this relation in the second term of Eq.~(\ref{mom1}) and making the shift $p\to p + uq$ we obtain\footnote{Before making the shift we also add linear terms proportional to $p^\mu$ and $p^\nu$ to the integrant of Eq. (\ref{mom1}), which by symmetry provide merely trivial contribution.},
\begin{eqnarray}
&&\int \mathcal{D}y~ \dot{y}^\mu_1 \dot{y}^\nu_2~ e^{iq y_1} e^{-iqy_2}  e^{ S_B(y) } =  \int \frac{d^D p}{(2\pi)^D} e^{-T [ (1-u)p^2 + u(p+q)^2 ]} \Big\{ \frac{2g^{\mu\nu}}{T}\delta(u) - (2p^\mu + q^\mu)(2p^\nu + q^\nu) \Big\}\,.
\label{mom2}
\end{eqnarray}
Finally, we need to integrate Eq.~(\ref{mom2}) over $T$ and the proper-time variables $\tau$. From the form of Eq.~(\ref{mom2}), it is easy to predict the structure of the result: the integration over $T$ yields a momentum denominator of the Feynman diagrams where the proper time variables play a role of Feynman parameters, as previously observed in Refs.~\cite{Bern:1991aq, Strassler:1992zr}. The delta function in Eq.~(\ref{mom2}) originates from the second derivative of the bosonic propagator and provides diagrams where interaction points of two incoming particles on the 
world-line coincide. Performing the integration, we get the following result
\begin{eqnarray}
&&\int^\infty_0 \frac{dT}{T} e^{-m^2T} \int^T_0 \prod^2_{i=1} d\tau_i \int \mathcal{D}y~ \dot{y}^\mu_1 \dot{y}^\nu_2~ e^{iq y_1} e^{-iqy_2}  e^{ S_B(y) } 
\nonumber\\
&&= (-i)^2\int \frac{d^D p}{(2\pi)^D} \frac{  (2p^\mu + q^\mu)(2p^\nu + q^\nu) - g^{\mu\nu}((p+q)^2 + m^2) - g^{\mu\nu} (p^2 + m^2) }{(p^2 + m^2)((p+q)^2 + m^2)}\,.
\label{mom3}
\end{eqnarray}
In the numerator of Eq.~(\ref{mom3}) one immediately  recognizes  the interaction vertices of scalar QED. 
\begin{figure}[htb]
 \begin{center}
 \includegraphics[width=120mm]{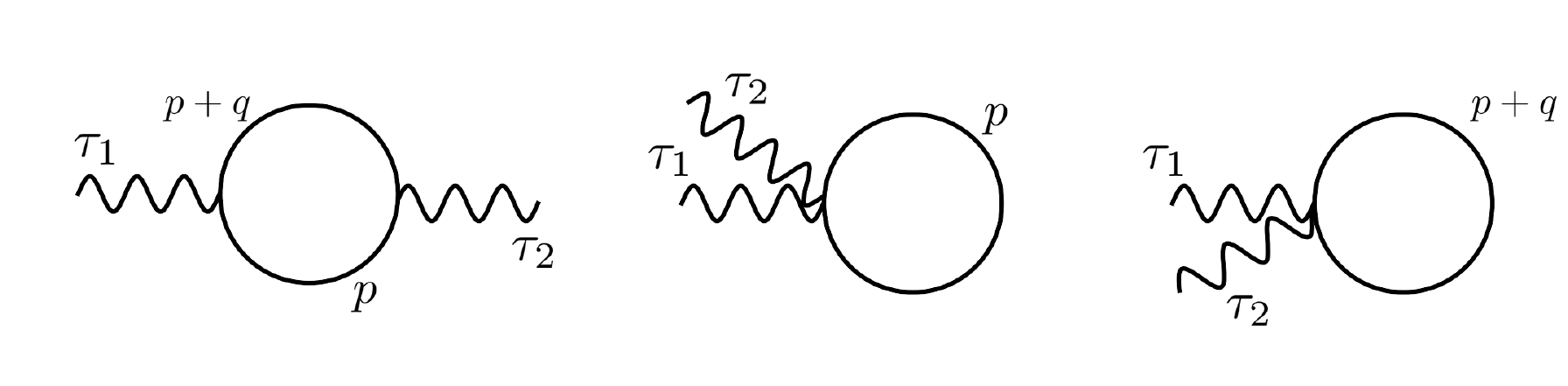}
 \end{center}
 \caption{\label{fig:5}Different contributions to the momentum representation of the world-line functional integral in Eq. (\ref{mom3})}
 \end{figure}

From Eq. (\ref{mom3}), one can derive a rule to reconstruct the form of the momentum integral with different numbers of incoming particles. For a given ordering of interactions\footnote{There is only one possible ordering in the case of two incoming particles},\footnote{That of course should be done in such a way that a chosen ordering of interactions on the world-line is not violated. By doing that we avoid the subtle point of how one should divide a contribution of the delta function in the derivative of the bosonic propagator between terms with different proper-time ordering.}  of the world-line with incoming particles, one should set a momentum flow and then write down an integral similar to Eq.~(\ref{mom3}), where the numerator has a product of factors $2p^\mu + q^\mu$ corresponding to the $\dot{y}^\mu$ currents in the functional integral, plus a sum of all possible combinations $\sim -g^{\mu\nu}$ where any two interaction vertices on the world-line coincide--see the delta function in Eq. (\ref{mom2})). The result in Eq.~(\ref{mom3}) is schematically presented in Fig. \ref{fig:5}.

\begin{figure}[htb]
 \begin{center}
 \includegraphics[width=160mm]{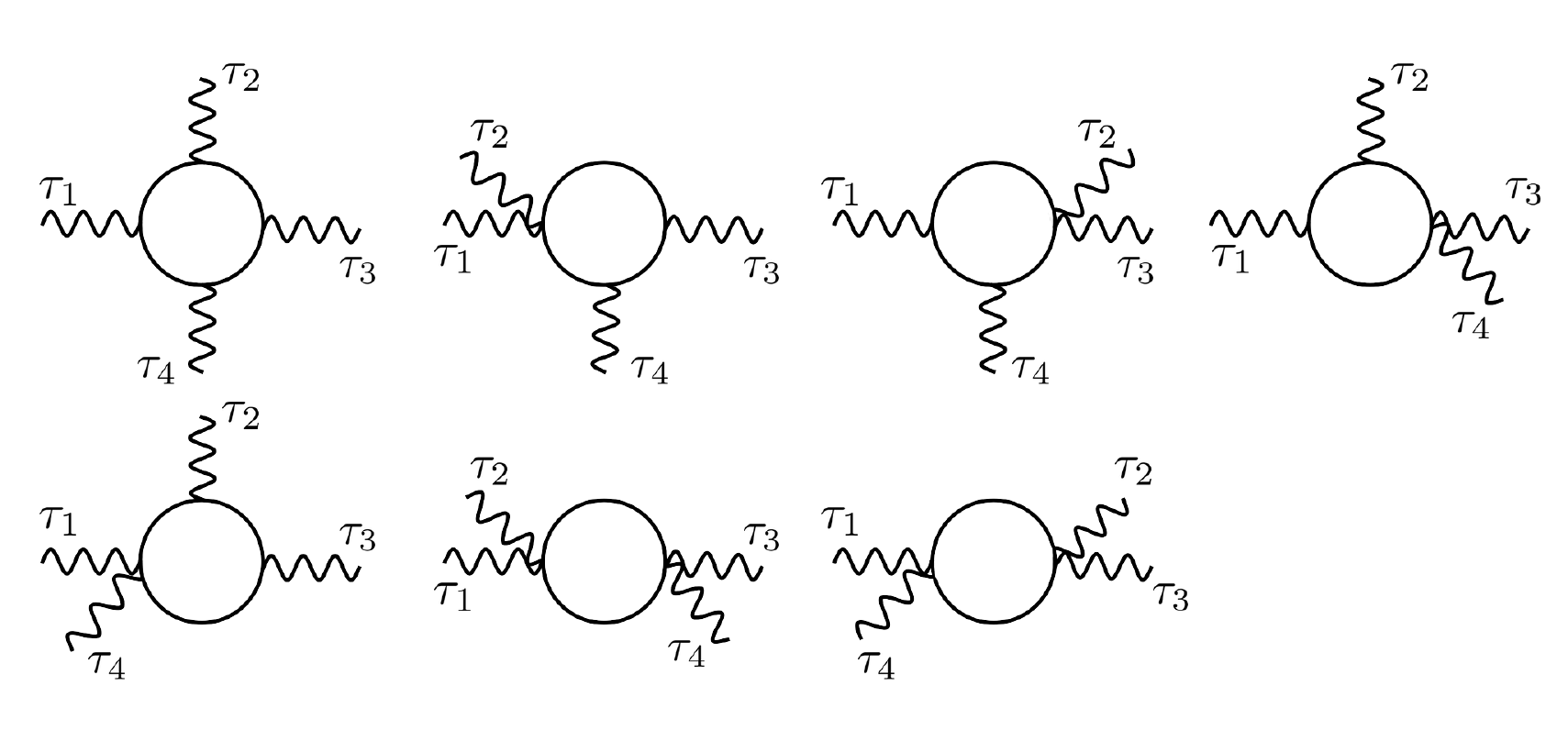}
 \end{center}
 \caption{\label{fig:6}Different contributions to the momentum representation of the world-line functional integral in eq. (\ref{fourc})}
 \end{figure}
 
Following this rule, one can easily write down momentum representation for an integral with arbitrary number of currents $\dot{x}^\mu$. For example, it yields the the following relations which we will use in the main part of the paper along with the Eq.~(\ref{mom3}):
\begin{eqnarray}
&&\int^\infty_0 \frac{dT}{T} e^{-m^2T} \int^T_0 \prod^2_{i=1} d\tau_i \int \mathcal{D}y\ \dot{y}^\mu_1~ e^{iq y_1}  e^{-iq y_2}  e^{S_B(y) }  =   - i  \int \frac{d^D p}{(2\pi)^D} \frac{2 p^\mu + q^\mu}{(p^2 + m^2) ((p + q)^2 + m^2) } 
\end{eqnarray}
and the more complicated identity,
\begin{eqnarray}
&&\int^\infty_0 \frac{dT}{T} e^{-m^2T} \int^T_0 \prod^4_{i=1} d\tau_i \int \mathcal{D}y~ \dot{y}^\mu_1 \dot{y}^\xi_2 \dot{y}^\nu_3 \dot{y}^\eta_4~ e^{iq y_1} e^{ik y_2} e^{-iqy_3}  e^{-ik y_4} e^{ S_B(y) } \Big|_{\tau_1 < \tau_2 < \tau_3 < \tau_4}
\nonumber\\
&&= (-i)^4\int \frac{d^D p}{(2\pi)^D} \frac{ \mathcal{N}^{\mu\xi\nu\eta}(p,q) }{(p^2 + m^2)((p+q)^2 + m^2)((p+q + k)^2 + m^2)((p+k)^2 + m^2)}
\label{fourc}
\end{eqnarray}
where the structure of the numerator can be easily understood from Fig. \ref{fig:6} to be 
\begin{eqnarray}
&&\mathcal{N}^{\mu\xi\nu\eta}(p,q) = \big(2p+q\big)^\mu\big(2p+2q + k\big)^\xi\big(2p + q + 2k\big)^\nu\big(2p+k\big)^\eta 
\nonumber\\
&&- g^{\mu\xi}\big((p+q)^2 + m^2\big)\big(2p + q + 2k\big)^\nu\big(2p+k\big)^\eta - g^{\xi\nu}\big((p+q + k)^2 + m^2\big)\big(2p+q\big)^\mu\big(2p+k\big)^\eta
\nonumber\\
&&- g^{\nu\eta}\big((p+k)^2 + m^2\big)\big(2p+q\big)^\mu\big(2p+2q + k\big)^\xi - g^{\mu\eta}\big(p^2 + m^2\big)\big(2p+2q + k\big)^\xi\big(2p + q + 2k\big)^\nu
\nonumber\\
&&+ g^{\mu\xi}g^{\eta\nu}\big((p+q)^2 + m^2\big)\big((p + k)^2 + m^2\big) + g^{\mu\eta}g^{\xi\nu}\big(p^2 + m^2\big)\big((p+q + k)^2 + m^2\big)
\label{fourcnum}
\end{eqnarray}
Expressions with other time orderings can be easily obtained from the above with an appropriate change of variables.

\section{World-lines and Feynman diagrams
\label{app:5}}

The world-line approach provides an alternative to the standard Feynman diagram description of a particle moving along the loop in an external background field. In the Feynman diagram technique, we deal with the functional integrals over quantum fields. These can be rewritten in terms of propagators and the spinor nature of the quark is described in terms of algebra of gamma matrixes. In the world-line approach, the perspective is completely different; the particle is characterized in terms of the world-line trajectory in coordinate space and the spinor nature of quark fields is described by a world-line trajectory in the space of Grassmann variables.

Despite the fact that the two formalisms are very different from one another, they describe the same physical object--a loop formed by the particle in a background fields. It is therefore very instructive to understand how the two methods are connected to each other and to develop a dictionary between the two approaches. For a general analysis of this problem, see Refs. \cite{Hostler:1985vb, Bern:1991an, Morgan:1995te, Mondragon:1995ab}.

 Let us consider the simple example of the vacuum polarization diagram in Fig. \ref{fig:2}c. We will show that structure of the trace of gamma matrixes for this diagram can be rewritten in terms of vertices of scalar QED and the spinor dependent interaction $\sim\sigma^{\mu\nu}F_{\mu\nu}$. We will demonstrate that scalar QED interactions correspond to the scalar current $\dot{x}$ in the world-line approach, while the spinor dependent interaction can be associated with the Grassmann current $\psi^\mu\psi^\nu F_{\mu\nu}$, see Eq. (\ref{EAQEDWl}).

There is a one-to-one correspondence between the two methods and one can immediately translate expressions obtained in one formalism into the language of the other. In our discussion we will reveal several subtle points of the world-line approach that are important for practical calculations. To the best of our knowledge, these have not been  fully addressed in literature.

We begin with the spinor trace in the vacuum polarization diagram,
\begin{eqnarray}
Tr\Big\{(\slashed{p}+\slashed{q} + m) \gamma^\mu (\slashed{p} + m)\gamma^\nu\Big\}
\label{polarFeynminit}
\end{eqnarray}
where $q$ is external momentum and $p$ is momentum of the loop; the factors in parentheses correspond to spinor contributions to the diagram. 
To rewrite this trace in terms of world-line structures, one should move the numerators of the quark propagators through the adjacent gamma matrices using the identity,
\begin{eqnarray}
&&(\slashed{p}+\slashed{q} + m)\gamma^\mu = \big\{(2p^\mu + q^\mu) + i \sigma^{\mu \rho} q_\rho\big\} - \gamma^\mu(\slashed{p} - m)\,,
\label{property}
\end{eqnarray}
which then gives 
\begin{eqnarray}
&&Tr\Big\{(\slashed{p}+\slashed{q} + m) \gamma^\mu (\slashed{p} + m)\gamma^\nu\Big\} =  Tr\Big\{\big((2p^\mu + q^\mu) + i \sigma^{\mu \rho} q_\rho\big) \big((2p^\nu + q^\nu)
 - i\sigma^{\nu\eta} q_\eta\big) \Big\}
\nonumber\\
&&- Tr\Big\{ (g^{\mu\nu} - i\sigma^{\mu\nu})(p^2 - m^2) \Big\} - Tr\Big\{ \big((2p^\mu + q^\mu) + i \sigma^{\mu \rho} q_\rho\big) \gamma^\nu(\slashed{p} + \slashed{q} - m)\Big\}\,.
\label{TrMove}
\end{eqnarray}

Now using the cyclic property of the trace, and some algebra, one can rewrite Eq.~(\ref{polarFeynminit}) as 
\begin{eqnarray}
&&Tr\Big\{(\slashed{p}+\slashed{q} + m) \gamma^\mu (\slashed{p} + m)\gamma^\nu\Big\} =  \frac{1}{2}Tr\Big\{ \big((2p^\mu + q^\mu) + i \sigma^{\mu \rho} q_\rho\big) \big((2p^\nu + q^\nu) - i\sigma^{\nu\eta} q_\eta\big)\Big\} 
\nonumber\\
&&-\frac{1}{2}Tr\Big\{ \big(g^{\mu\nu} - i\sigma^{\mu\nu}\big) \Big\} \big((p+q)^2 - m^2\big) - \frac{1}{2}Tr\Big\{\big(g^{\mu\nu} - i\sigma^{\mu\nu}\big)\Big\} \big(p^2 - m^2\big)
\label{TrRes}
\end{eqnarray}
We should note here that while the step of the derivation when we apply the cyclic property of trace may seem like a trivial manipulation, it explicitly shows that the technique we discuss here is uniquely applicable to loop diagrams.  In the world-line approach, this is realized only when the world-line trajectory has the topology of a closed loop. One otherwise has to include boundary effects from the ends of the open world-line trajectory. In the Feynman diagram approach, these will correspond to external spinors. 

In other words, while it seems natural to break the world-line and write down a representation for a spinning  particle moving between two finite points in space, one should be very careful with this  because boundary effects absent in the world-line action in Eq.~(\ref{EAQEDWl}) have to be properly taken into account. Thus while tempting, the generalization of the world-line representation to the case of propagators of spinning particles, i.e. open world-lines, is not straightforward~\cite{Fradkin:1991ci,Fainberg:1987jr}.

As we discussed previously, the structure of the polarization tensor in Eq.~(\ref{polarization}) in the main text embeds the product of two world-line currents of the form 
\begin{eqnarray}
j^\mu = \dot{x}^\mu + i\psi^\mu\psi^\rho q_\rho\,.
\label{current2}
\end{eqnarray}
Indeed the $2p^\mu + q^\mu$ term in Eq. (\ref{TrRes}) is just the interaction vertex of scalar QED. See also the first term in Eq. (\ref{mom3}) which corresponds to the bosonic world-line contribution in Fig. \ref{fig:5}a.

The second term of Eq. (\ref{current2}) can be easily associated with the structures  $\sigma^{\mu\rho}q_\rho$ in Eq. (\ref{TrRes}). The simplest way to understand this is to compare the Grassmann functional integral of $\psi^\mu(\tau)\psi^\rho(\tau) q_\rho$ with the trace of $\sigma^{\mu\rho}q_\rho$.\footnote{Note that both Grassmann variables here depend on the same proper time $\tau$, otherwise each of them should be associated with a gamma matrix, i.e. $\gamma^\mu \sim \psi^\mu(\tau)$.} For example, compare Eq. (\ref{GrFInt3}) and
\begin{eqnarray}
&&\frac{1}{4}Tr\{ \sigma^{\mu_1\nu_1} \sigma^{\mu_2\nu_2} \} 
 = g^{\mu_1\mu_2}g^{\nu_1\nu_2} - g^{\mu_1\nu_2}g^{\nu_1\mu_2}\,.
\end{eqnarray}
One can exploit effectively this equivalence between the functional integral over $\psi^\mu\psi^\rho q_\rho$ and trace of $\sigma^{\mu\rho}q_\rho$. Indeed, while the calculation of Grassmann functional integrals is quite involved, see Eq. (\ref{GrAfterExp2}), and not yet realized in computer codes, there are plenty of tools for the computation of traces of gamma matrices.

Now let us consider the second line of Eq. (\ref{TrRes}). There is a term $\sim g^{\mu\nu}(p^2 - m^2)$ which we have already seen in calculation of the bosonic world-line functional integral summarized in Eq. (\ref{mom3}). The origin of this term is $\delta(\tau_1 - \tau_2)$ in the derivative of the bosonic world-line propagator, as seen in  Eq. (\ref{wlpropDer2}). It corresponds to Fig. \ref{fig:5}c where two scalar currents $\dot{x}$ meet each other on the world-line. In terms of the Feynman rules of scalar QED, it is of course the seagull interaction.

There is also a term $\sim \sigma^{\mu\nu}$ in the second line of Eq. (\ref{TrRes}) whose contribution to the vacuum polarization trace in Eq.~(\ref{polarFeynminit}) is zero. However if we construct the  representation in Eq.~(\ref{TrRes}) for more complicated Feynman diagrams (for example, with more than two external photons), it is not necessarily the case. While this on the surface suggests a discrepancy between the world-line formalism and that of Feynman diagrams, this discrepancy vanishes if we consider a sum of Feynman diagrams with all possible insertions of external photons. In terms of world-lines, this corresponds to the integration over all possible $\tau_i$, as Eq. (\ref{WModFu3}). In this case, there is a cancelation of $\sigma^{\mu\nu}$  terms from different diagrams amongst each other.  This comparison of the two techniques therefore suggests that  $\sim \sigma^{\mu\nu}$ terms with particular proper time orderings must be treated with care in world-line computations. This will be especially important when one explores polarized parton distributions.

\bibliography{wlines}

\end{document}